\documentclass[%
 reprint,
nofootinbib,
 amsmath,amssymb,
 aps,
]{revtex4-2}

\usepackage{aas_macros}
\usepackage{graphicx}% Include figure files
\usepackage{dcolumn}% Align table columns on decimal point
\usepackage{bm}% bold math
\usepackage{hyperref} %for refer table of contents
\usepackage{cleveref}
\hypersetup{
    colorlinks,
    citecolor= red,
    filecolor=red,
    linkcolor= blue,
    urlcolor= blue
}

\usepackage{Zallman,lettrine}

\begin{document}

% \preprint{APS/123-QED}

\title{Galactic and extragalactic probe of dark matter with LISA's binary black holes}% Force line breaks with \\
\thanks{sohanghodla9@gmail.com}%

\author{Sohan Ghodla}
\affiliation{Department of Physics and Astronomy, Colgate University, 13 Oak Dr, Hamilton, 13346 NY, USA \\ Department of Physics, University of Auckland, Private Bag 92019, Auckland, New Zealand}

\date{\today}% It is always \today, today,
              % but any date may be explicitly specified

\begin{abstract}

The upcoming LISA mission will be able to detect gravitational waves from galactic and extragalactic compact binaries. Here, we report on LISA's capability to probe dark matter around these binaries if the latter constitute black holes. By analyzing the variation in the chirp mass of the binary, we show that depending on the black hole masses, LISA should be able to probe their surrounding dark matter to a luminosity distance of $\approx 1$ Gpc if such binaries are located within the inner $\approx 10$ pc of their galactic center for particle-like dark matter or near the galactic solitonic core for wave-like dark matter. However, for the latter, the density of dark matter near the galactic center must be higher than predicted from dark matter only simulations. Even if a null result is recorded during the course of observation of \textit{well-localized} binaries, one can rule out certain parameter spaces of dark matter as being the dominant contributor to the matter budget of the Universe.

\end{abstract}

%\keywords{Suggested keywords}%Use showkeys class option if keyword
                              %display desired
\maketitle

%\tableofcontents

\section{Introduction}

\lettrine[lines=3]{W}{}ith the emergence of gravitational wave astronomy, the use of gravitational radiation as a probe of dark matter has gained prominence, with black holes being one of the prime targets \cite{Eda:2013gg, Macedo:2013qea, Kavanagh:2020cfn, Dai:2023cft, Traykova:2023qyv, Akil:2023kym, Kadota:2023wlm, Mitra:2023sny, Cai:2023ykr, Califano:2024xzt}. 
Like all astrophysical bodies, black holes reside within a dark matter surrounding and would slowly accumulate this matter over time. 
However, in contrast to dark matter accumulation in celestial bodies, such as planets \cite{Bramante:2019fhi, Mack:Beacom:2007, leane2021exoplanets}, stars \cite{Bell:2011sn, km3netsun, In:2017kcf, Gould:1987ju, Press:1985ug, Peter:2009mk, Garani:2017jcj, Kouvaris:2015nsa, Croon:2024}, white dwarfs \cite{Leung:2013pra, Graham:2018efk, Acevedo:2019gre, graham2018white},  neutron stars \cite{Goldman:1989, Gould:1990, Kouvaris:2007ay, Garani:2019fpa, Kouvaris_2010, Bramante:2014zca, Gresham:2018rqo, Bramante:2017ulk, de_neutron_2010} which have to rely on as of yet uncertain scattering processes, the presence of event horizon makes black holes unique sinks for their surrounding material. 
This renders the process of dark matter capture by black holes immune to the uncertain nature of dark matter's interactions with standard-model particles.

A direct consequence of a steady dark matter accumulation is a continuous increase in the black hole's mass. Thus, if the black hole is part of a binary system, this mass growth leads to a faster orbital inspiral - driven by the emission of gravitational radiation - compared to the case when the binary lives in a vacuum. 
Within the first order of post-Newtonian expansion, the gravitational waveform frequency evolution depends only on the binary's chirp mass $\mathcal{M}$.
To this end, by accurately measuring both the gravitational wave frequency and its rate of change — see Eq.~\eqref{eq: chirp mass} later — one can calculate the change in $\mathcal{M}$. However, detecting such subtle variations requires a prolonged period of observation.

Current ground-based gravitational wave detectors operating in the $\approx 10-10^3$ Hz frequency range observe stellar-mass binaries only for a few seconds before merger \cite{Abbott_O3a_2021, Abbott_O3b_2021}. In contrast, the upcoming space-based LISA mission will open a new window in the 0.1-100 mHz frequency regime \cite{LISA:2022yao}, with planned observation times of 4-10 years, offering an excellent resolution into mass variation. Over this extended duration, LISA will be capable of observing binaries across a mass range spanning stellar to supermassive black holes (SMBHs). Moreover, for strain amplitudes with a large signal-to-noise ratio (SNR), many gravitational wave-emitting systems will be individually resolvable \cite{Robson:Cornish:2019}. As a consequence, stellar-mass binary black holes (BBHs) present within the Milky Way will provide us with the opportunity to probe their surroundings. On the other hand, for extragalactic distances, intermediate-mass black holes (IMBHs) with masses $10^2 - 10^4 M_\odot$ become particularly important.

In this work, we investigate the impact of a dark matter environment on the evolution of BBHs that radiate within the LISA frequency band and are individually resolvable (as a binary). 
 We show that certain BBHs that reside near the center of their galaxy will strongly interact with their surrounding dark matter halo. Depending on the distance to the source, for those with gravitational wave frequencies $f \gtrsim 1$ mHz, changes in their chirp mass can be inferred from the observed frequency evolution, allowing us to probe their environment.
For the latter, we focus on two variants of dark matter, namely particle-like dark matter and (wave-like) scalar-field dark matter. 
We also discuss the possible impact of the presence of interstellar gas around the BBHs on our analysis.  

It is anticipated that within its 4-10 yr of operation, LISA will detect $\mathcal{O}(10)-\mathcal{O}(100)$ BBHs distributed across the Milky Way\footnote{There is also a possibility of dynamical inband formation, which is not covered by the aforementioned study.} \cite{Wagg:2021cst}. 
Although it is difficult to make such estimates for the rate of IMBH encounters with their host SMBH, IMBHs will be detectable up to redshift two \cite{Jani:2019ffg}. Depending on the masses of these black holes, we demonstrate that LISA can probe the surrounding dark matter of BBHs to a luminosity distance of $\approx 1$ Gpc provided these binaries are located either within the inner $\approx 10$ pc of their galactic center (for particle-like dark matter) or near their galactic solitonic core (for wave-like dark matter). However, for the latter, the density of dark matter near the galactic center must be higher than predicted from dark matter only simulations. We show that a null result during the course of observation of spatially well-localized binaries could still rule out certain parameter spaces of dark matter as being the dominant contributor to the matter budget of the Universe.

The remainder of this paper is organized as follows. In Section~\ref{sec: capture of dark matter by black holes}, we discuss the formalism employed for investigating the capture of dark matter by black holes. In Section~\ref{sec: Dynamics of a BBH embedded in a dark matter surrounding}, we study the dynamics of a BBH immersed in such a dark matter environment and then discuss  its observability in Section~\ref{sec: Observability}. In Section~\ref{sec: results}, we present the result for galactic and extragalactic probes of dark matter, followed by a brief discussion in Section~\ref{sec: discussion}.

\vspace{-10pt}
\section{Capture of dark matter by black holes} \label{sec: capture of dark matter by black holes}

For the following investigation, we assume that the dark matter medium is at rest w.r.t. the galactic rest frame but has a non-zero dispersion velocity. We consider two variants for dark matter, namely those that have masses $\gtrsim 1$ eV and can be treated as point particle-like and those that have masses $\ll 1$ eV resulting in wave-like properties that can be modeled via scalar fields. While the former has been successful on large scales within the $\Lambda$CDM paradigm, the latter is also naturally (i.e., independent of baryonic physics) able to account for the small-scale discrepancies  \cite{Schive:2014}  typically associated with particle-like dark matter \cite{Klypin:1999uc, deBlok:2010}.

\vspace{-12pt}
\subsection{Particle-like dark matter} \label{sec: particle-like DM}

For a Schwarzschild black hole with mass $m$ that is moving at a supersonic velocity $v_\infty$ w.r.t. the asymptotic/unperturbed background medium with density $\rho_\infty$, matter within the impact parameter $r_a = {2Gm} / {v_\infty^2}$ will be captured by the black hole. The resulting accretion cross-section can be estimated as $\sigma_{\rm HL} = \pi r_a^2$ which results in the Hoyle-Lyttleton mass accretion rate \cite{hoyle1941accretion}
\begin{equation}
    \dot m_{\rm HL} = \sigma_{\rm HL} \rho_\infty v_\infty = \frac{4 \pi G^2 m^2 \rho_{\infty}}{v_{\infty}^3} \,.
    \label{eq: Hoyle Lyttlenton}
\end{equation}
Meanwhile, following Bondi \cite{Bondi1952}, for a black hole that has $v_\infty = 0$,  the mass accretion rate reads 
\begin{equation}
    \dot m_{\rm B}  = \frac{4 \pi \lambda G^{2} m^{2} \rho_\infty}{c_{s}^3} \,; \quad \lambda = \frac{1}{4}\left(\frac{2}{5-3 \gamma}\right)^{\frac{5-3 \gamma}{2(\gamma-1)}} \,,
    \label{eq: Bondi}
\end{equation}
where $c_s = c \sqrt{\gamma \Theta_\infty}$ is the local sound speed and 
\begin{equation}
    \Theta_\infty = \frac{k_{B} T}{m_\chi c^{2}}\,,
\end{equation}
is the dark matter dimensionless temperature. Above, $T, m_\chi, k_B, c$ are the dark matter temperature, particle mass, Boltzmann constant, and the speed of light in vacuum, respectively. 

In deriving Eq.~\eqref{eq: Bondi}, it is assumed that the background medium can be treated as a polytrope with the equation of state $P \propto \rho^\gamma$, where $\gamma \leq 5/3$ for a non-relativistic fluid, and that dark matter can be treated as a non-relativistic medium. A relativistic extension of the Bondi formalism was later conducted by Michel in Ref. \cite{michel1972accretion} (for a review, see Appendix of Ref. \cite{Tejeda:2020}). However, for a medium with $\Theta_\infty \lesssim 10^{-5}$, the mass accretion rate can be well modelled by Eq.~\eqref{eq: Bondi} (cf., \citep{aguayo2021spherical}). Since the current upper bound on the dark matter's dimensionless temperature is expected to be $\Theta_\infty \approx \mathcal{O}(10^{-8})$ \citep{Avelino_DM_soundspeed}, thus one can consider Eq.~\eqref{eq: Bondi} to hold for dark matter accretion irrespective of it being cold or warm.  

Since $\dot m_{\rm HL} \propto v_\infty^{-3}$, to generate a larger accretion rate, for the current work, we are interested in the scenario where the black hole either moves at a subsonic or marginally supersonic velocity. 
To this end, using Eq.~\eqref{eq: Hoyle Lyttlenton} and \eqref{eq: Bondi},  a simple interpolation gives us the Bondi-Hoyle \cite{Bond_Hoyle_1944} mass accretion rate, 
\begin{equation}
   \dot m_{\rm BH}  = \frac{4 \pi \lambda G^{2} m^{2} \rho_\infty}{\left(c_{s}^{2} + v^{2}_\infty \right)^{3 / 2}} \,,
   \label{eq: Bondi-Hoyle accretion}
\end{equation}
which holds for both the subsonic and supersonic scenarios. 
The above expression for the accretion rate would be weakly sensitive to the spin of the black holes since the accretion cross-section is set at a large distance from the black hole. Thus, the calculations performed would not be significantly impacted by the assumption of Schwarzschild geometry.

To determine $\rho_\infty$, we use an NFW profile for the dark matter halo; see Eq.~\eqref{eq: NFW profile}. In the inner region, the halo can experience adiabatic compression due to the presence of the SMBH, resulting in a spike in dark matter density \cite{Gondolo:1999ef, Davies_2020, Kim:2022mdj}. To give conservative estimates for $\dot m$, we disregard the presence of such a density spike in this work.

\vspace{-20pt}
\subsection{Scalar field dark matter}

If dark matter is ultralight, its wavelength $\lambda_{\rm DM} \gg 2Gm$. This is in contrast to particle-like dark matter discussed in Section~\ref{sec: particle-like DM}, where $\lambda_{\rm DM} \ll 2Gm$, thus requiring an alternative approach to calculate its capture rate.
Here we assume that such a dark matter candidate can be modeled via a massive scalar field living on a curved background. The treatment of such long wavelength scalar field absorption by a Schwarzschild black hole is provided in \cite{Unruh_1976}. The resulting mass accretion rate reads
\begin{equation}
     \dot m_{\rm S} = \sigma_{S}(m, m_\chi, v_\infty) \rho_\infty v_\infty \,,
     \label{eq: Unruh accretion}
\end{equation}
where (with $\hbar = c = 1$)
\begin{equation}
    \begin{aligned}
        \sigma_{\rm S} = \frac{16 \pi G^{2} m^{2}}{v_\infty} \frac{\xi}{1-e^{-\xi}} ; \quad   \xi=2 \pi G m m_\chi \frac{1+v^{2}_\infty}{v_\infty \sqrt{1-v^{2}_\infty}} \,. 
    \end{aligned}
    \label{eq: Unruh cross-section}
\end{equation}

The capture rate of such long-wavelength scalar fields is expected to be relatively weak compared to point-particle dark matter and likely to be important when the black hole is present in dense dark matter environments, such as near the center of our galaxy. A generic prediction for such ultralight fields is that they will produce a solitonic core near the galactic center (e.g., \cite{Schive:2014, hui2017ultralight}). In the absence of self-interaction, the soliton is purely supported via quantum pressure, and its density can be approximated as \cite{Schive:2014} 
\begin{equation}
    \rho=\rho_c\left(1 + \tilde \lambda\left(\frac r{r_c}\right)^2\right)^{-8} \,,
\end{equation}
where $\tilde \lambda = 2^{1/8} - 1$.
Additionally, the core's central density can be written as \cite{hui2017ultralight}
\begin{equation}
    \rho_{c}= \rho_{0} \left(\frac{G m_\chi^{2}}{c^2  \hbar^{2}}\right)^{3} M_{\rm sol}^{4}  \,,
    \label{eq: core density}
\end{equation}
where $M_{\rm sol}$ is the mass of the soliton and $\rho_0 = 0.00440$ \cite{hui2017ultralight}. The core radius $r_c$ is defined as $\rho(r_c) = \rho_c / 2$ and takes the form \cite{Schive:2014}
\begin{equation}
    r_c=\left(\frac{\rho_c}{0.019 {M}_\odot\ \text{pc}^{-3}}\right)^{-1/4}\left(\frac{m_\chi}{10^{-22}\text{eV}}\right)^{-1/2}\text{kpc }\,.
    \label{eq: core radius}
\end{equation}

The virial velocity of the solitonic core takes the form
\begin{equation}
     v_\infty \equiv v_{\mathrm{vir}} = \frac{G M_{\rm sol} m_\chi}{\hbar} w_{0}^{1 / 2}
     \label{eq: virial velocity}
\end{equation}
with $w_0 = 0.10851$ \cite{hui2017ultralight}.
Since our target black holes will be going in a circular orbit around their galactic center with velocity $v_c$, with $v_c \ll v_{\mathrm{vir}}$ ($v_c$ calculated in Appendix~\ref{sec: galactic circular velocity}), thus, we make the assumption that $v_\infty \equiv v_{\mathrm{vir}}$, which would otherwise be determined by $v_c$. 

The masses of black holes considered here, along with the expression for $v_\infty$ in Eq.~\eqref{eq: virial velocity}, suggests $\xi \ll 1$, implying that Eq.~\eqref{eq: Unruh accretion} can be reduced to
\begin{equation}
   \frac{\mathrm{d} m_{\rm S}}{\mathrm{d} t}  = \frac{16\pi(G m)^2 \rho_\infty}{c^3} \,.
   \label{eq: soliton mass accretion rate}
\end{equation}
Substituting the density of the soliton in Eq.~\eqref{eq: soliton mass accretion rate} thus results in the scalar field accretion rate 
\begin{equation}
\begin{aligned}
    \frac{\mathrm{d} m_{\rm S}}{\mathrm{d} t} = \frac{2.5 \, M_{\odot}}{10^{3} \, \mathrm{yr}} & \left(\frac{m}{10^{9} M_{\odot}}\right)^{2}\left(\frac{m_\chi}{10^{-22} \mathrm{eV}}\right)^{6}\left(\frac{M_{\rm sol}}{10^{10} M_{\odot}}\right)^{4} \\
    & \times \left(1+\lambda\left(\frac{r}{r_c}\right)^2\right)^{-8} \,.
    \label{eq: mass accretion rate for scalar field}
\end{aligned}
\end{equation}
Above, $M_{\rm sol}$ can be calculated using the soliton - halo mass relation acquired from numerical simulations as \cite{Davies_2020}
\begin{equation}
    M_{\rm sol} =1.25\times10^{9}\left(\frac{M_{{\mathrm{halo}}}}{10^{12} {~M}_{\odot}}\right)^{1/3}\left(\frac{m_\chi}{10^{-22}\mathrm{~eV}}\right)^{-1} {~M}_{\odot} \,,
    \label{eq: soliton halo mass relation}
\end{equation}
where $M_{\rm halo}$ is the mass of the dark matter halo. This implies that $M_{\rm sol} \propto m_\chi^{-1}$ and, as we later discuss, for Milky Way like galaxy with $M_{\rm halo} \approx 10^{12} M_\odot$ results in a negligible impact on $\mathcal{M}$ of the binary.
To this end, for the sake of demonstration of the formalism discussed here, we also consider an alternative scenario where we fix the Milky Way's soliton mass to $M_{\rm sol} = 10^{9} M_\odot$ (the latter value is based on Ref. \cite{de2020dynamical}) irrespective of $m_\chi$.

\section{Dynamics of a BBH immersed in a dark matter surrounding} \label{sec: Dynamics of a BBH embedded in a dark matter surrounding}

\subsection{Mass evolution}

Since LISA's binaries will have an orbital period of $10 - 10^4$~s \cite{LISA:2022yao}, this implies $r_a \gg a$, $a$ being the binary's semi-major axis. Thus far away from the binary, the infalling point-particle dark matter will perceive the binary as a single object with mass $M = m_1 + m_2$ \cite{Farris:2010, Kaaz:2019, Comerford:2019}. This gives the total Bondi-Hoyle accretion rate on the BBH as
\begin{equation}
     \dot{M}_{\rm BH}  =  \dot{m}_{1, \rm BH} + \dot{m}_{2, \rm BH} + 2 \delta \frac{4 \pi \lambda G^{2} m_1 m_2 \rho_\infty}{\left(c_{s}^{2} + v^{2}_\infty \right)^{3 / 2}} \,,
     \label{eq: Bondi-Hoyle on BBH}
\end{equation}
where $0 \leq \delta \leq 1$.

As matter falls into the total potential of mass $M$,  far away from the BBH, $\delta = 1$. Then, to maintain a steady-state mass accretion rate, this captured mass would eventually have to be accreted. The mass contained in the last term in Eq.~\eqref{eq: Bondi-Hoyle on BBH} would be distributed among the component masses. As the current treatment only serves as a test of LISA's capability to detect dark matter owing to variation in $M$, in the following, for simplicity, we present our results assuming $\delta = 0$, which provides a floor for the expected evolution of the BBH mass. E.g., for the special case of $m_1 = m_2$, $\dot M_{{\rm BH}, \delta = 1} = 2 \dot M_{{\rm BH}, \delta = 0}$ with $\dot M_{\rm BH}$ lying between the two values for other values of $\delta$.

A similar effect would manifest during the descent of wave-like dark matter on the binary's collective gravitational potential, cf.  Eq.~\eqref{eq: soliton mass accretion rate}. Nevertheless, as previously, we treat the infall and absorption of dark matter on each black hole separately for the latter case as well.

\vspace{-12pt}
\subsection{Orbital decay}

\subsubsection{Due to emission of gravitational radiation}

The rate of orbital decay of a BBH is governed by the rate at which it emits gravitational radiation. For the most part of the evolution, the orbital decay of BBHs considered here can be modeled in the weak field regime. If averaged over the orbital period, the decay takes the form \cite{peters1964gravitational}
\begin{equation}
    \left. \frac{d a}{d t} \right|_{\rm GW} \hspace{-6pt} = -\frac{64}{5} \frac{G^3 \mu M^2}{c^5 a^3\left(1-e^2\right)^{7 / 2}}\left(1+\frac{73}{24} e^2+\frac{37}{96} e^4\right) + \mathcal{O}(\dot m),
    \label{eq: GW orbital decay rate}
\end{equation}
where $\mu = m_1m_2/(m_1 + m_2)$ is the reduced mass of the binary and $e$ is the eccentricity. We note that due to the capture of dark matter, the above masses are time-varying and only the leading-order effect has been considered. Similarly, the eccentricity of the binary also changes as
\begin{equation}
    \left. \frac{d e}{d t} \right|_{\rm GW} \hspace{-6pt} = - \frac{304}{15} e \frac{G^3 \mu M^2}{c^5 a^4\left(1-e^2\right)^{5 / 2}}\left(1+\frac{121}{304} e^2\right) + \mathcal{O}(\dot m) \,,
    \label{eq: GW eccentricity decay rate}
\end{equation}
where the subscript ``GW'' indicates that the decay is mediated by gravitational wave emission.

\vspace{-15pt}
\subsubsection{Due to impact on orbital angular momentum}

We are interested in BBHs with orbital frequency $\in$ [$10^{-4}, 0.1$] Hz that only enter the merger phase (i.e., strong curvature regime) after spending at least $\mathcal{O}({\rm yr})$ in the LISA band. At such frequencies, the orbital velocity of the binary component $v_{\rm orb} \gg v_{\infty}$ but at the same time for the most part, $v_{\rm orb} \ll c$ such that a fully relativistic treatment can be ignored. As $v_{\rm orb} \gg v_{\infty}$, regardless of the direction of motion of the BBH w.r.t. the background dark matter medium, \textit{averaged over one full orbit}, this implies that dark matter accretion would contribute negligibly to the orbital angular momentum $L$ of the binary. Thus, we assume that outside of the loss in gravitational wave emission,
\begin{equation}
    L=\sqrt{G \mu^{2} M a\left(1-e^{2}\right)}
\end{equation}
is a constant of the motion. For a mass-changing binary, this results in orbital decay at rate
\begin{equation}
    \left. \frac{da}{dt} \right|_L = - a \left[\frac{ d}{dt} \ln{ (\mu^2 M}) - \frac{2e \dot{e}}{1 - e^2}\right]  \,,
    \label{eq: AM orbital decay rate}
\end{equation}
and if averaged over one period, following equation (25) in Ref. \cite{hadjidemetriou1963two}, we find that the eccentricity to evolve as
\begin{equation}
    \left. \frac{de}{dt} \right|_L= - \frac{e \dot M}{M} \,.
    \label{eq: AM eccentricity decay rate}
\end{equation}
Thus, the net orbit decay rate can be written as $\dot a =  \dot a |_{\rm GW} +  \dot a |_L$ and the net eccentricity decay rate as $\dot e =  \dot e |_{\rm GW} +  \dot e |_L$. We note that for LISA's BBHs both Eqs.~(\ref{eq: AM orbital decay rate}, \ref{eq: AM eccentricity decay rate}) have a sub-leading impact on the BBH's orbital dynamics compared to Eqs.~(\ref{eq: GW orbital decay rate}, \ref{eq: GW eccentricity decay rate}).

\vspace{-5pt}
\subsection{Gravitational wave frequency evolution}

For BBHs orbiting in a Keplerian orbit, the orbital frequency $f_{\rm orb}$ satisfies
\begin{equation}
    f_{\rm orb}^2 = \frac{GM}{4 \pi^2 a^3} \,.
\end{equation}
Thus, if the binary accretes dark matter, the evolution of $f_{\rm orb}$ can be calculated as
\begin{equation}
    \frac{df_{\rm orb}}{dt} = \frac{G^{1/2}}{4 \pi} a^{-3/2} M^{-1/2} \left[ \dot M - \frac{3M}{a} \dot a \right] \,,
    \label{eq: orbital frequency evolutin}
\end{equation}
where in the scenario when the BBH lives in a vacuum, we can set $\dot M = 0$. From Eq.~\eqref{eq: orbital frequency evolutin}, one can then calculate the gravitational wave frequency evolution, noting that the gravitational wave frequency in the $n$th harmonic should be given by $f = n f_{\rm orb}$.

\vspace{-5pt}
\section{Observability} \label{sec: Observability}

\subsection{An evolving chirp mass}

To restate, we are interested in determining the changes in the evolution of the BBH properties, owing to the capture of surrounding dark matter. The leading-order post-Newtonian orbital evolution of a gravitational wave emitting binary is determined by its chirp mass $\mathcal{M} = \mu^{3/5} M^{2/5}$ and in the case where $\dot M = 0$, $\mathcal{M}$ remains invariant over time. Observationally, for certain binaries, LISA may be able to measure their gravitational wave frequency $f$ as well as the frequency drift $\dot f$ yielding
\begin{equation}
    \mathcal{M} = \frac{c^{3}}{G}\left(\frac{5}{96} \pi^{-8 / 3} f^{-11 / 3} \dot{f}\right)^{3 / 5}\,,
    \label{eq: chirp mass}
\end{equation}
which LISA can measure with the resolution \cite{Takahashi:Seto:2002}
\begin{equation}
   \sigma_{\mathcal{M}}  \approx \frac{3}{5} \frac{\mathcal{M}}{\dot{f}} \sigma_{\dot{f}} \,.
\end{equation}
Moreover, for circular binaries, $\dot f$ can be measured with the resolution
\begin{equation}
    \sigma_{\dot{f}} \approx 0.43\left(\frac{\bar\rho_2}{10}\right)^{-1} T_{\rm obs}^{-2} \,,
\end{equation}
where $\bar \rho_2$ is the SNR of the gravitational wave of the binary in the second orbital harmonic (i.e., $n=2$) and is defined later in Eq.~\eqref{eq: SNR}. 
This yields
\begin{equation}
     \sigma_{\mathcal{M}}  \approx 0.258 \left(\frac{\bar\rho_2}{10}\right)^{-1} \frac{ \mathcal{M}}{\dot{f}} T_{\rm obs}^{-2} \,,
\end{equation}
implying that a longer observation time will lead to a smaller error in the measured value of $\mathcal{M}$. Assuming that we measure $\mathcal{M}$ for two different observation times $t_1, t_2$ (with $t_2 > t_1$), we can then calculate the change in the binary's chirp mass over the observation time as $\Delta \mathcal{M} = \mathcal{M}(t_2) - \mathcal{M}(t_1)$, where the change in $\mathcal{M}$ is driven by the capture of dark matter. For a successful detection of this change, we require that the net measurement uncertainty remain smaller than the measured change in $\mathcal{M}$, i.e., 
\begin{equation}
    \Delta \mathcal{M}(t_2) > \sigma_{\mathcal{M}(t_1)} + \sigma_{\mathcal{M}(t_2)}\,.
\end{equation}
where $\sigma_{\mathcal{M}}(t_1) > \sigma_{\mathcal{M}}(t_2)$. For calculational simplicity, in the current work, we adopt a more stringent criterion, i.e., 
\begin{equation}
    \Delta \mathcal{M}(t > t_1) \geq 2 \sigma_{\mathcal{M}(t_1)}
\end{equation}
for a successful detection.

\vspace{-10pt}
\subsection{Strain calculation}

For a given value of $f, \mathcal{M}, e$ and source luminosity distance $D_L$,  the dimensionless gravitational wave strain amplitude of the binary in the $n=2$ orbital
harmonic (i.e., the most dominant harmonic at low eccentricities) is \cite{Nelemans:2001hp, Robson:Cornish:2019}
\begin{equation}
  h_{2}=\frac{8 G^{5 / 3} \mathcal{M}^{5 / 3} \pi^{2 / 3} f^{2 / 3}}{5^{1 / 2} D_L c^{4}}\left[1-\frac{5}{2} e^{2}+\frac{35}{24} e^{4}+O\left(e^{6}\right)\right] \,.
\end{equation}
In calculating $h_2$ for a given binary, we will treat $D_L$ to remain constant over the course of 
observation. Moreover, since we are not interested in a particular target, here we consider the gravitational wave SNR to be averaged over inclination, sky location, and gravitational wave polarization over an observing time $T_{\rm obs}$. This yields \cite{Robson:Cornish:2019}
\begin{equation}
    \bar{\rho_2}=\frac{h_2}{S_{n}(f)^{1 / 2}} \sqrt{T_{\rm obs}} \,.
    \label{eq: SNR}
\end{equation}
where $S_n(f)$ is LISA's noise power spectral density and is defined in Appendix~\ref{sec: LISA sensitivity}.
For higher harmonics, $h_n \propto \sqrt{g(n,e)} / n$ and become increasingly important for $e \gtrsim 0.2$ (see figure 3 in Ref. \cite{Peters:Mathews:1963}, the expression for $g(n,e)$ can be also be found therein). 
For simplicity below, we present our calculation only for the case $e = 0$ for which the binary radiates at $n = 2$. As the intensity of the emitted radiation is directly proportional to $e$, so our results can be treated as the floor of LISA's sensitivity.

\vspace{-5pt}
\section{Result} \label{sec: results}

\begin{figure}
    \centering
    \includegraphics[width=1\linewidth]{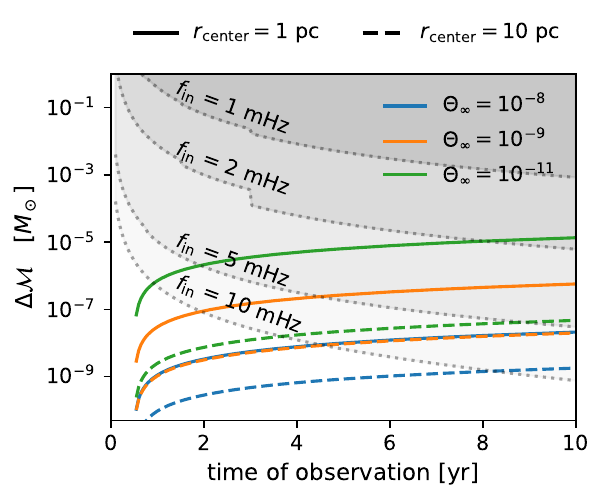}
    % \vspace{-20pt}
    \caption{Variation in the chirp mass $\mathcal{M}$ of a BBH with $m_1, m_2 = 25M_\odot$ and $e = 0$ that is located at a distance of $r_{\rm center} = 1$ pc, 10 pc from the center of Milky Way. Colored lines represent $\Theta_\infty$ - the dimensionless temperature of particle-like dark matter. The shaded region above the dotted curves represents the region where variations in $\mathcal{M}$ of a BBH with initial gravitational wave frequency $f_{\rm in}$ (i.e., $f$ at the beginning of the observation run) are detectable by LISA.}
    \label{fig: galactic_point_particle_DM_detectability}
\end{figure}

\begin{figure}
    \centering
    \includegraphics[width=1\linewidth]{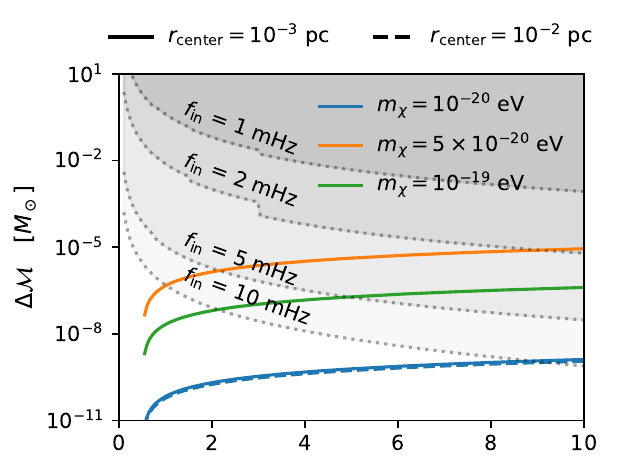}
    \includegraphics[width=1\linewidth]{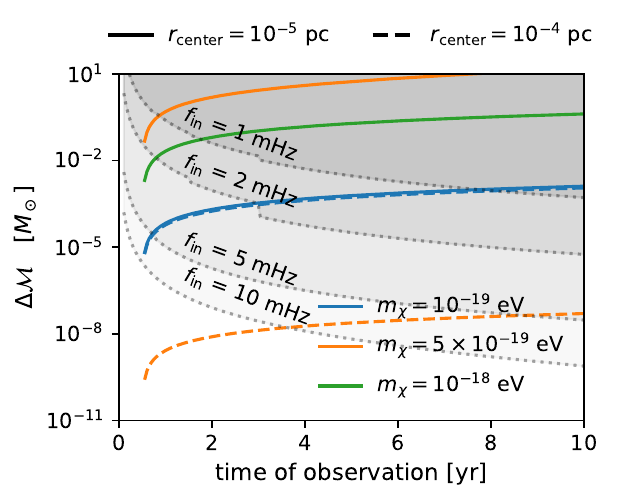}
    % \vspace{-20pt}
    \caption{Same as Fig. \ref{fig: galactic_point_particle_DM_detectability} but now for wave-like dark matter. The colors of the curves represent the mass of dark matter, and the BBH is assumed to be much closer to the galactic center. For reference, the Schwarzschild radius of Sgr A$^*$ is $\approx 4 \times 10^{-7}$ pc.}
    \label{fig: galactic_scalar_field_DM_detectability}
\end{figure}

\subsection{Galactic sources} \label{sec: results galactic}

Given the inverse dependence of $h_2$ on $D_L$, binaries within the galaxy will be the optimal target.
As mentioned earlier, we assume such binaries to be in a circular orbit around the galactic center with velocity $v_c$, which will be the source of $v_\infty$ in Eq.~\eqref{eq: Bondi-Hoyle accretion}.
In addition to a large $M$, Eq.~\eqref{eq: Bondi-Hoyle accretion} and \eqref{eq: soliton mass accretion rate} suggests that a larger change in $\mathcal{M}$ requires $\rho_\infty$ to be large and $v_c$ to be small. Thus, the ideal location to probe dark matter using LISA would be near the galactic center.

The calculation of $v_c$ for black holes residing in the Milky Way's disk is presented in Appendix~\ref{sec: galactic circular velocity}. Since the role of the galactic bulge remains uncertain in the inner region, we assume that it is the NFW halo that determines the value of $v_c$ near the galactic center. Nevertheless, we compare $v_c$ resulting from the presence of an NFW halo and the bulge independently in Fig.~\ref{fig: rotation_curve}. 
As discussed in Section~\ref{sec: capture of dark matter by black holes}, in contrast to particle-like dark matter, absorption of ultralight scalar fields will only be important when the BBH resides within the galactic soliton. For such a case, $v_{\rm vir} \gg v_c$, thus the effect on a non-zero $v_c$ can be ignored.

Fig.~\ref{fig: galactic_point_particle_DM_detectability} shows the variation in $\mathcal{M}$ of a BBH with $m_1, m_2 = 25 M_\odot$ due to the presence of particle-like dark matter. The shaded area in the figure shows the region of detectability by LISA and is dependent on the emitted gravitational wave frequency. The change in $\mathcal{M}$ depends on the magnitude of $\Theta_\infty$ and $r_{\rm center}$ (the distance of the BBH from the galactic center) with smaller values resulting in more favorable outcomes. We note that the kink in the curve with initial frequency $f_{\rm in} = 2$ mHz is an artifact and results from the interpolation carried over the LISA sensitivity data presented in Table 1 of Ref. \cite{Robson:Cornish:2019}.

Similar to Fig.~\ref{fig: galactic_point_particle_DM_detectability}, Fig.~\ref{fig: galactic_scalar_field_DM_detectability} shows the variation in $\mathcal{M}$ for the case where the BBH is immersed in a wave-like dark matter surrounding of $M_{\rm sol} = 10^9 M_\odot$ with the corresponding dark matter masses shown in the figure legend. As the mass of dark matter increases, the density of the solitonic core increases. However, this also decreases the value of $r_c$ in Eq.~\ref{eq: core radius}. Thus, we place the BBH closer to the galactic center for a larger value of $m_\chi$.
For both Fig.~\ref{fig: galactic_point_particle_DM_detectability} and \ref{fig: galactic_scalar_field_DM_detectability}, we find that within 4-10 yr of operation, LISA will be able to probe the surroundings of the BBHs if such systems are detected during the observational run. 
However, $M_{\rm sol}$ is expected to be a dynamical quantity that changes as one varies $m_\chi$ \cite{Davies_2020}. For the sake of demonstration, let us pick $m_\chi = 10^{-19}$ eV as this mass results in a detectable effect in Fig.~\ref{fig: galactic_scalar_field_DM_detectability}. Then, based on Eq.~\eqref{eq: soliton halo mass relation}, the corresponding $M_{\rm sol} = 1.25 \times 10^6 M_\odot$. This reduces the mass accretion rate in Eq.~\eqref{eq: mass accretion rate for scalar field} by a factor of $\approx 10^{12}$ compared to the case where $M_{\rm sol} = 10^9 M_\odot$ and gets worse as $m_\chi$ increases. As the value of $M_{\rm sol}$ has a direct impact on $\rho_c$ in Eq.~\eqref{eq: core density}, adopting a larger value of $M_{\rm sol}$ results in a larger $\rho$. This effect is akin to the impact of adiabatic compression of the soliton on dark matter density, where the compression is driven by the gravitational influence of the SMBH at the galactic center \cite{Davies_2020, Kim:2022mdj}.
Thus, we conclude that wave-like dark matter is unlikely to cause any noticeable variation in $\mathcal{M}$ unless the soliton gets adiabatically compressed to result in larger values of $\rho$ near the galactic center.

Until now, we assumed the BBH mass ratio $q \equiv m_2 / m_1 = 1$. To investigate the impact of a lower value of $q$, we reproduce Fig.~\ref{fig: galactic_point_particle_DM_detectability} and \ref{fig: galactic_scalar_field_DM_detectability} in Fig.~\ref{fig: galactic detectability with q = 0.1} in Appendix~\ref{sec: additional figures}, assuming $q = 0.1$. As anticipated, a lower mass ratio reduces the prospects of detectability, meaning for smaller $q$ values, a higher value of $M$ would be desirable to probe the surrounding dark matter.

\begin{figure*}
    \centering
    \includegraphics[width=0.825\linewidth]{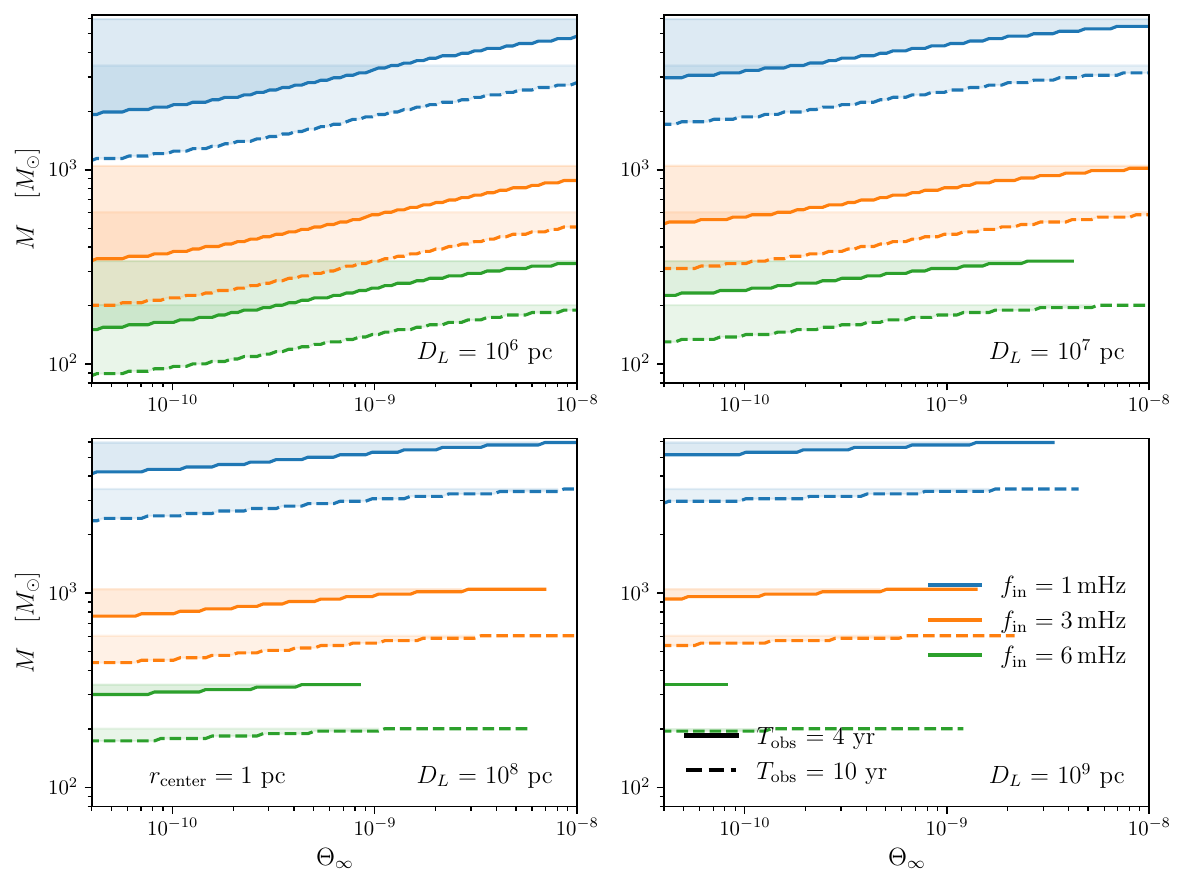}
     \includegraphics[width=0.825\linewidth]{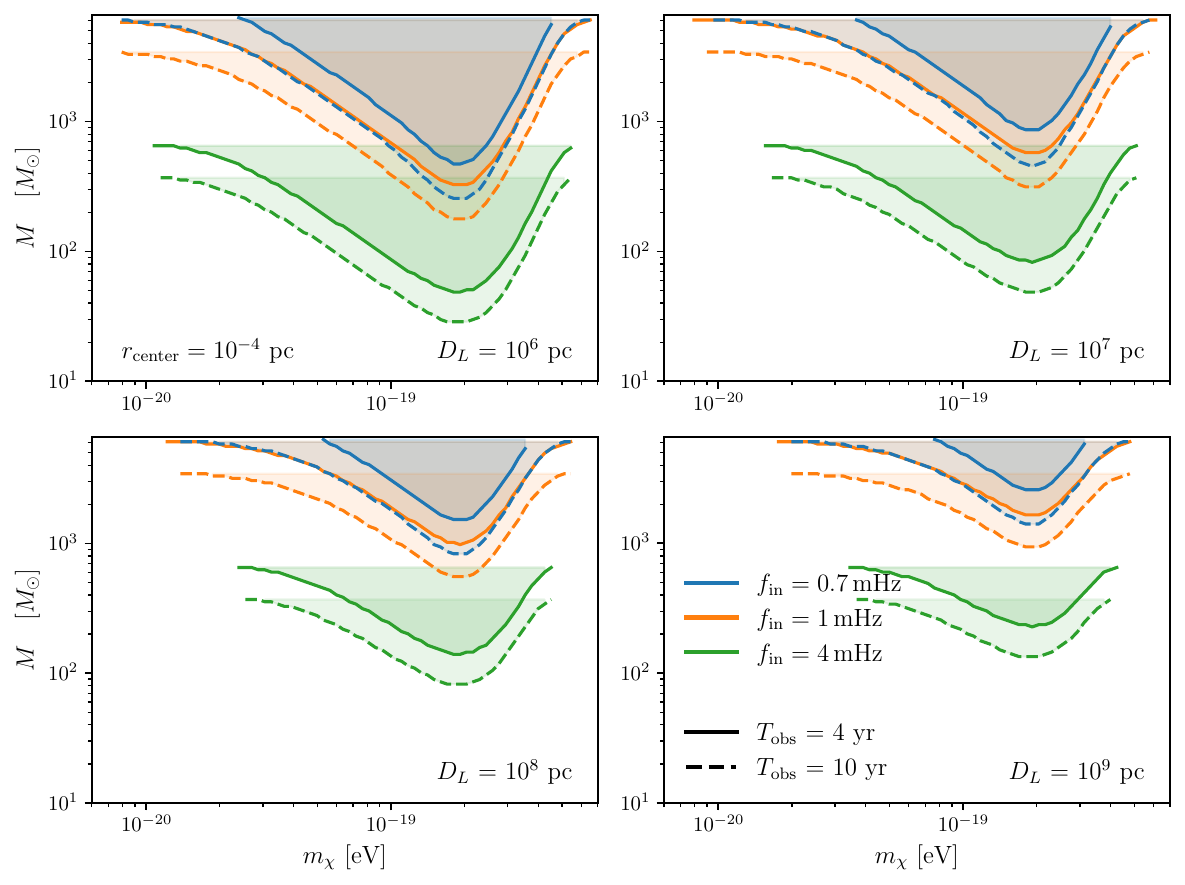}
     \vspace{-14pt}
    \caption{\textit{Top two rows}: Shaded region represents the parameter space where variations in $\mathcal{M}$ of a BBH (with total mass $M$ and $q = 1$) are detectable by LISA. Colors represent $f_{\rm in}$ of the BBH at the onset of observation. The two linestyles represent the length of observation. Various regions are clipped from above as these massive BBHs merge before the end of observation. The BBHs are assumed to be located at $r_{\rm center} = 1$ pc from their galactic center. The various values of $D_L$ are also shown near the lower left of each plot. \textit{Bottom two rows}: Same as top two rows but representing wave-like dark matter with $r_{\rm center} = 10^{-4}$ pc.}
    \label{fig: extragalactic detectability}
\end{figure*}

\subsection{Extragalactic sources}

LISA will be sensitive to IMBHs out to redshift two \cite{Jani:2019ffg}. Thus, in this section, we apply the above-discussed methodology to larger values of $D_L$.
For simplicity, we take the Milky Way's dark matter profile as a prototype for all galaxies at extragalactic distances.

The resulting extragalactic detectability prospects are shown in Fig.~\ref{fig: extragalactic detectability} and \ref{fig: extragalactic detectability with q = 0.1} (in Appendix~\ref{sec: additional figures}) for $q = 1, 0.1$ respectively, where we again set $M_{\rm sol} = 10^9 M_\odot$. For demonstration purposes, here we have chosen a subset of $f_{\rm in}$ values and assumed a \emph{fixed observation time} of $T_{\rm obs} = 4, 10$ yr. At the end of $T_{\rm obs}$, $\Delta \mathcal{M}$ is calculated as $\Delta \mathcal{M} (T_{\rm obs}) = \mathcal{M}(T_{\rm obs}) - \mathcal{M}(T_{\rm obs} - 0.1 {\rm yr})$. Many BBHs merge within this time span and are thus removed from the figure, as is evident from the truncated region at the top of the shaded area \footnote{In principle, one could fill other regions of Fig.~\ref{fig: extragalactic detectability} and \ref{fig: extragalactic detectability with q = 0.1} by calculating $\Delta \mathcal{M}$ as one continuously varies $T_{\rm obs}$.}.

 These figures show that observing $\mathcal{M}$ variation could allow us to probe the dimensionless temperature in range $\Theta_\infty \lesssim 10^{-8}$ for the case of particle dark matter\footnote{The maximum value of $\Theta_\infty$ is estimated to be $\mathcal{O}(10^{-8})$ \cite{Avelino_DM_soundspeed}.} and $m_\chi \in [10^{-20} , 5 \times 10^{-19}]$ eV for wave-like dark matter up to a distance of $\approx 1$ Gpc. The latter depends on the value of $r_{\rm center}$, and one can imagine that the detection prospects would be more favorable for smaller values of $r_{\rm center}$.

\vspace{-20pt}
\section{Discussion} \label{sec: discussion}

\vspace{-10pt}
\subsection{Impact of a baryonic surrounding} \label{sec: impact of ISM}

Here, we explore the impact of a baryonic surrounding on the evolution of black hole mass and discuss if such a phenomenon could erase any imprint of the surrounding dark matter. Solving Eq.~\eqref{eq: Bondi-Hoyle accretion} yields the mass evolution of the black hole as 
\begin{equation}
    m(t) =  \eta m_i \,; \quad  \eta(t) = - \frac{1}{m_i \alpha t - m_i \alpha t_i -1}; \quad \eta(t) \geq 1 \,,
    \label{eq: mass evolution}
\end{equation}
where $m_i$ is the mass of the black hole at some initial time $t_i$ and $\alpha = \frac{4 \pi \lambda G^2 \rho_\infty}{(c_s^2 + v_\infty^2)^{3/2}}$.
The accretion of the interstellar medium (ISM) will be most efficient when the black hole comoves w.r.t. to its surrounding medium. Thus, for the case of baryonic matter accretion, we set $v_\infty = 0$ in Eq.~\eqref{eq: mass evolution}. Moreover, the value of $\rho_\infty$ for the baryonic environment will depend on the nature of ISM in the vicinity of the black hole. A favorable scenario occurs when the ISM is cold and hence contains neutral particles, typically hydrogen. Such a medium is referred to as a cold neutral medium (CNM) and contains an average particle number density $n_H  \approx (20 -50)$ cm$^{-3}$ and a temperature $T \approx 100$ K \cite{Draine:2011}.

\begin{figure*}
    \centering
    \includegraphics[width=1\linewidth]{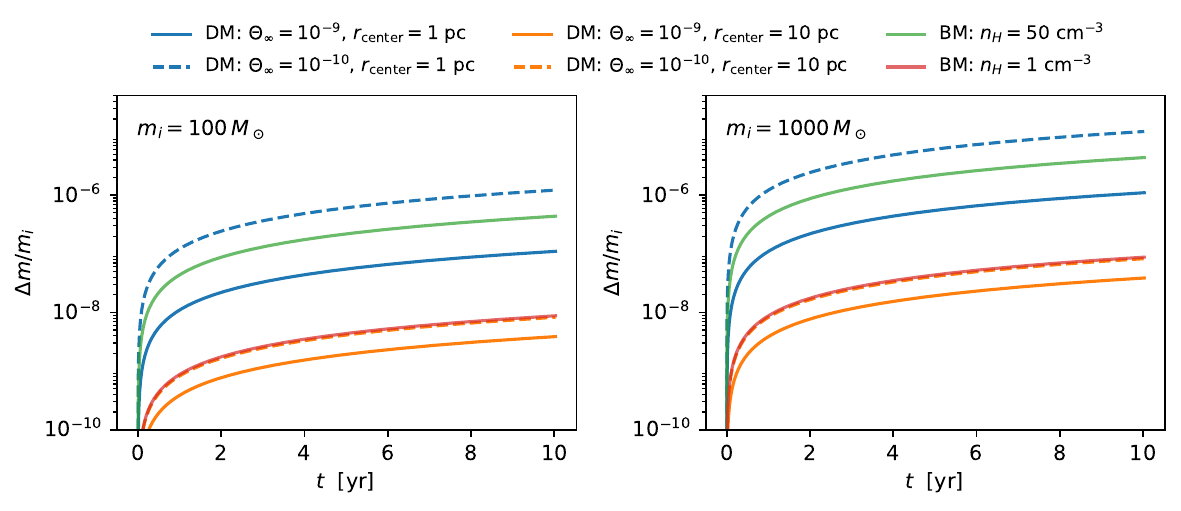}
    \vspace{-20pt}
    \caption{The fractional change in a black hole's mass with initial mass $m_i$ located at $r_{\rm center} = $ 1 pc, 10 pc from the galactic center. The LHS (RHS) figure shows $m_i = 100 M_\odot$ ($m_i = 1000 M_\odot$). In the legend, ``DM'' and ``BM'' stand for dark matter and baryonic matter, respectively. $\Theta_\infty$ represents dark matter dimensionless temperature and $n_H$ represents baryonic particle number density. Although we have considered only two values for $m_i$, a similar trend holds for other values as well.}
    \label{fig: DM_vs_BM_accretion}
\end{figure*}

It is likely that most of the surrounding CNM would have already been accreted by the time the BBH enters the LISA frequency band. Nevertheless, it is worthwhile to investigate how the black hole mass evolution compares to the scenario when the mass growth is supported by dark matter and baryonic matter accretion, respectively. To this end, Fig.~\ref{fig: DM_vs_BM_accretion} shows the mass evolution of black holes with $m_i = 100 M_\odot, \, 1000 M_\odot$ immersed in a baryonic and particle-like dark matter medium with variable properties.
Here, the ISM is considered to be cold with $T = 100$~K.

However, there is a possibility that baryonic matter may form an accretion disk around the binary (by borrowing angular momentum from the BBH's orbit) and thus heat up the surrounding gas. This will cause the temperature to rise, and for $n_H \gtrsim 1$ cm$^{-3}$, the baryonic matter will have a subdominant impact on the black hole mass evolution once $T \gtrsim 200$ K. 
To be quantitative, let us assume that only a fraction $\nu$ of the accreted mass is assimilated into an accretion disk (with the rest getting accreted directly). Then, one finds that the accretion luminosity w.r.t. the Eddington luminosity $f_L$ should evolve as \cite{Ghodla:2024gda}
%  for the parameter values of $n_H = 1$ cm$^{-3}$ and $T_\infty = 100$ K,
% 
\begin{equation}
   f_L = \epsilon \frac{\dot{m}}{ \Dot{m}_{\rm Edd}} = 3.9 \times 10^{-3}  \epsilon  \nu \left( \frac{n_H}{ {\rm cm}^3} \right)  \left(\frac{m}{M_{\odot}}\right) \left( \frac{100 \; \rm{K}}{T} \right)^{3/2} \,.
    \label{mass accretion w.r.t Eddintion rate}
\end{equation}
For illustrative purposes, assuming $\nu = 0.1$, $T = 100$~K,  $m \in [10 M_\odot, 1000 M_\odot]$ and the accretion efficiency $\epsilon = 1/16$, yields $f_L \in (0.00024 n_H, 0.024 n_H)$. Noting that for $f_L \gtrsim 0.001$, one excepts the accretion disk to radiate as a Shakura-cSunyaev accretion flow \cite{Shakura_Sunyaev_1973, Yuan_Narayan_2014}, thus, for black holes with $m \gtrsim 100 M_\odot$, the effect of feedback could be very strong (even for $n_H = 1$ cm$^{-3}$), likely shutting off accretion of baryonic ISM due to a rise in its temperature.

\vspace{-10pt}
\subsection{Implication of a null result and the issue of localization}

Regardless of the presence of a baryonic environment near the black holes, LISA not detecting any $\mathcal{M}$ evolution during the course of observation of optimal targets would imply that the underlying dark-matter model used for performing the calculation here is not the dominant form of dark matter around the BBHs. While it can still be a subdominant component, it has to be below the threshold at which $\mathcal{M}$ evolution can be observed. Depending on the value of $M, q$, and $D_L$, the method proposed here can be used to perform such a check for a large parameter space of dark matter.

However, LISA alone will be unable to localize most BBHs to a subgalactic scale if observed at larger luminosity distances. Following equation 15 in Ref.~\cite{Takahashi:Seto:2002} suggests that of the BBH mass range considered here, only the massive IMBHs within $D_L \lesssim 1 $Mpc will have sufficient angular resolution to be localized to an $\approx 10$ pc scale. Thus, a null result will only be useful for such values of $D_L$. This is because, as we showed, BBHs at larger distances from their galactic center would not interact strongly with their surrounding dark matter. Thus, for them, one may not expect to see any appreciable $\mathcal{M}$ variation anyway. 

There remains a possibility that if the BBH is located close to its galactic center - i.e., within $\lesssim$ $\mathcal{O}(100)$ Schwarzschild radii of the SMBH, the SMBH-BBH system might radiate gravitational waves such that the inspiral of the BBH into the SMBH also becomes resolvable by LISA. This will result in two distinct (overlapping) gravitational wave signals, both arising from the same location in space but from two different sources. This can be a strong clue that the BBH signal may arise from the center of the host galaxy hosting the SMBH. In such a scenario, the null result mentioned above becomes a useful tool to much larger values of $D_L$.

\vspace{-12pt}
\subsection{Conclusion}

In this work, we have attempted to estimate the floor of the possible $\mathcal{M}$ variation caused by dark matter accretion onto LISA's BBHs and whether such a variation is detectable. In particular, we set $\delta = 0$ - Eq.~\eqref{eq: Bondi-Hoyle on BBH}, assumed circular orbits, i.e.,  $e= 0$, and considered no adiabatic compression of the dark matter halo \cite{Gondolo:1999ef, Kim:2022mdj}, the latter leading to larger dark matter density near the galactic center (although see Section~\ref{sec: results galactic}).

Our analysis demonstrates that the BBH surrounding environment can be probed by measuring variation in the binary's chirp mass over the course of LISA's observation. As the impact of baryonic matter accretion may be subdominant for IMBHs (Section~\ref{sec: impact of ISM}), any significant variation in their $\mathcal{M}$ value can be attributed to the presence of dark matter.
These variations depend on the intrinsic properties of dark matter, allowing us to probe them up to a luminosity distance of approximately 1 Gpc, depending on the BBH’s location relative to the galactic center and the particle physics property of dark matter.
For the case where the BBH is very close to the galactic SMBH,  the presence of an AGN disk can contaminate the resulting gravitational wave signal. It remains to be seen if meaningful information about the surrounding dark matter can still be recovered under such a scenario. 

The code used for generating the figures in this paper is available at \url{https://github.com/SohanGhodla/Probing-dark-matter-with-LISA}.

\vspace{-12pt}
\begin{acknowledgments}
This work was in part supported by the University of Auckland Doctoral Scholarship and the Picker Interdisciplinary Science Institute at Colgate University.
\end{acknowledgments}

\appendix

\vspace{-10pt}
\section{Galactic circular velocity} \label{sec: galactic circular velocity}

% To restate, our focus will remain on dark matter capture by BBHs going in a circular orbit around their galactic center. Both Eq.~\eqref{eq: Bondi-Hoyle accretion} and \eqref{eq: mass accretion rate for scalar field} depend on the value of $v_\infty$.  As the dark matter medium is expected to be roughly at rest w.r.t. the galactic rest from, the circular velocity of these binaries will be the source of $v_\infty$. Since dark matter density falls off with radial distance and the relative motion between dark matter and the BBH increases with radial distance, the prospects of studying dark matter are stronger if such a binary is detected close to the galactic center. Nevertheless, as discussed in the main text, for the capture of ultralight scalar fields, $v_{\rm vir} \gg v_c$. Thus, for them, the effect of the relative motion can be ignored.

\begin{figure}
    \centering
    \includegraphics[width = 1\linewidth]{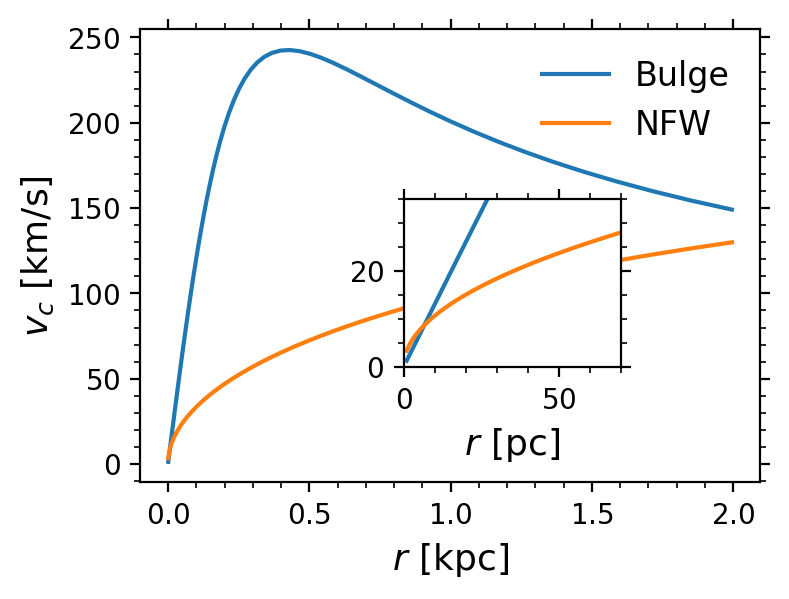}
    \vspace{-10pt}
    \caption{Galactic circular velocity of Milky Way under the scenario when $\Phi$ is dominated by the galactic bulge (blue) and the NFW halo (orange), respectively. The inset figure shows the region near the center in units of parsec.}
    \label{fig: rotation_curve}
\end{figure}

We calculate the galactic mid-plane ($z=0$) circular velocity as a function of distance $r$ from the galactic center as 
\begin{equation}
    v_{\mathrm{c}}^{2}(r)=\left.r \frac{\partial \Phi}{\partial r}\right|_{z=0} \,,
\end{equation}
where the gravitational potential $\Phi$ is the sum of contributions from different components, denoted as $\Phi_i$ below. To account for $\Phi_i$, we include contributions from a spherical dark matter halo and a spherical galactic bulge.  When a component is described by its density $\rho_i$ - e.g., Eq.~\eqref{eq: NFW profile}, we obtain the corresponding $\Phi_i$ through the Poisson equation
\begin{equation}
    \nabla^{2} \Phi_{i}=4 \pi G \rho_{i} \,.
\end{equation}
In the inner region where $v_c$ is small, $\Phi$ could be dominated by the contribution from the bulge. The potential for the latter can be approximated as a Plummer potential and takes the form
\begin{equation}
    \Phi_{\text {Plummer}}(r)=-\frac{G M_{\text {bulge }}}{\sqrt{r^{2}+r_{b}^{2}}} \,,
\end{equation}
where $M_{\rm bulge} = 1.067 \times 10^{10} M_\odot$ and $r_b = 0.3$ kpc is the cut-off radius \cite{Pouliasis_2017}. The resulting velocity takes the form 
\begin{equation}
    v_c^2 (r) = \frac{G M_{\rm bulge} r^2} { (r^2 + r_b^2)^{3/2} }
\end{equation}
and is shown in Fig.~\ref{fig: rotation_curve}.

Although $v_c$ for $r \in [5, 25]$ kpc has been well measured \cite{GAIA:2019}, uncertainties remain regarding its value in the inner region of the galaxy. 
Moreover, there is no unanimous agreement on the nature of the galactic bulge (for small $r$), with some studies suggesting that it might be subdominant \cite{reid2014trigonometric}. In such a case, it is the dark matter halo that dictates the value of $v_c$ in the inner region of the galaxy (which we assume to be the case in the present work). To calculate the resulting value of $v_c$, we adopt an NFW profile \cite{NFW_1997} for the Milky Way's dark matter halo as
\begin{equation}
    \rho_{\mathrm{NFW}}(r)= \rho_{0} \frac{r_{s}}{r} \left(1+\frac{r}{r_{s}}\right)^{-2} \,,
    \label{eq: NFW profile}
\end{equation}
where $\rho_0 = 0.052 \, M_\odot$pc$^{-3}$, $r_s = 8.1$ kpc are the normalisation constant and scale radius, respectively \cite{lin2019dark}. Eq.~\eqref{eq: NFW profile}, thus yields
\begin{equation}
    v_c^2(r) = 4 \pi G \rho_0 \frac{r_s^3}{r} \left[ \frac{r_s}{r_s + r} - \ln{ \left( \frac{r_s}{r_s + r} \right) - 1} \right] 
\end{equation}
and has been plotted in Fig.~\ref{fig: rotation_curve}.

\vspace{-5pt}
\section{LISA sensitivity} \label{sec: LISA sensitivity}

To determine the sensitivity of LISA towards the detection of a given binary, we follow the fitting relations provided in Ref. \cite{Robson:Cornish:2019}. The LISA noise power spectral density $S_n$ has a contribution from two sources
\begin{equation}
    S_{n}(f)=S_{n}^{\rm{ins}}(f)+S_{n}^{\rm{WDB}}(f) \,,
\end{equation}
where the former term results from the instrument noise while the latter results from the confusion noise from the population of unresolved galactic white-dwarf binaries.
Explicitly,
\begin{equation}
\begin{aligned}
    S_{n}^{\rm{ins}}(f)=& A_{1}\left(P_{\rm{OMS}}+2\left[1+\cos ^{2}\left(f / f_{\star}\right)\right] \frac{P_{\rm{acc}}}{(2 \pi f)^{4}}\right) \\
    & \times \left(1+\frac{6}{10} \frac{f^{2}}{f_{\star}^{2}}\right) \,,
\end{aligned}
\end{equation}
where
\begin{equation}
\begin{aligned}
    &A_{1} = \frac{10}{3 L^{2}}, \, L=2.5 {\rm Gm}, \, f_{\star}=19.09 \rm{mHz} \,, \\
    & P_{\rm{OMS}} = (1.5 \times 10^{-11} {\rm~m})^2  \left[1+\left(\frac{2 \rm{mHz}}{f}\right)^{4}\right] \rm{Hz}^{-1} \,, \\ & P_{\rm acc} = \left(3 \times 10^{-15} \rm{~ms}^{-2}\right)^{2}\left[1+\left(\frac{0.4 \rm{mHz}}{f} \right)^{2}\right] \\ 
    & \quad \quad \times {\left[1+\left(\frac{f}{8 \rm{mHz}}\right)^{4}\right] \rm{Hz}^{-1}} \,.
\end{aligned}
\end{equation}
Additionally, 
\begin{equation}
\begin{aligned}
    & S_{n}^{\rm{WDB}} = A_{2} f^{-7 / 3} e^{-f^{\alpha}+\beta f \sin (\kappa f)} \\ 
    & \quad \quad \quad \quad \quad   \times \left[1+\tanh \left(\gamma\left(f_{k}-f\right)\right)\right] \,\, \rm{Hz}^{-1} \,,
\end{aligned}
\end{equation}
where, depending on the magnitude of $T_{\rm obs}$ the parameters $A_{2}, \alpha, \beta, \kappa, \gamma, f_{k}$ have been interpolated from Table 1 of Ref. \cite{Robson:Cornish:2019}. 
% They do not provided data beyond 4 years.

\vspace{-20pt}
\section{Additional figures} \label{sec: additional figures}

\begin{figure}
    \centering
    \includegraphics[width=1\linewidth]{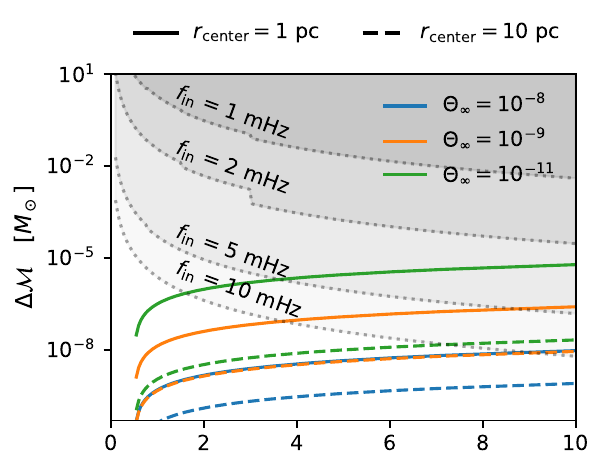}
    \includegraphics[width=1\linewidth]{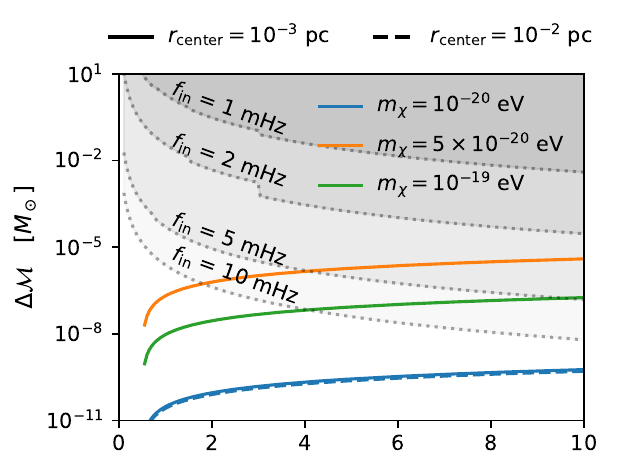}
    \includegraphics[width=1\linewidth]{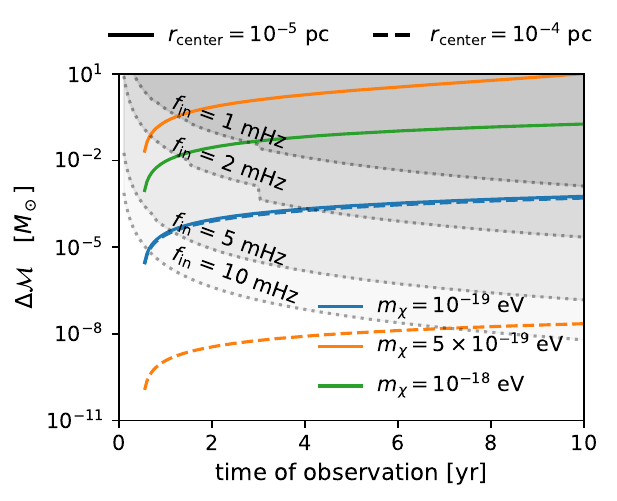}

    \caption{Same as Fig. \ref{fig: galactic_point_particle_DM_detectability} and \ref{fig: galactic_scalar_field_DM_detectability} but with $q = 0.1$  ($M = 50 M_\odot$).}
    \label{fig: galactic detectability with q = 0.1}
\end{figure}

\begin{figure*}
    \centering
    \includegraphics[width=0.85\linewidth]{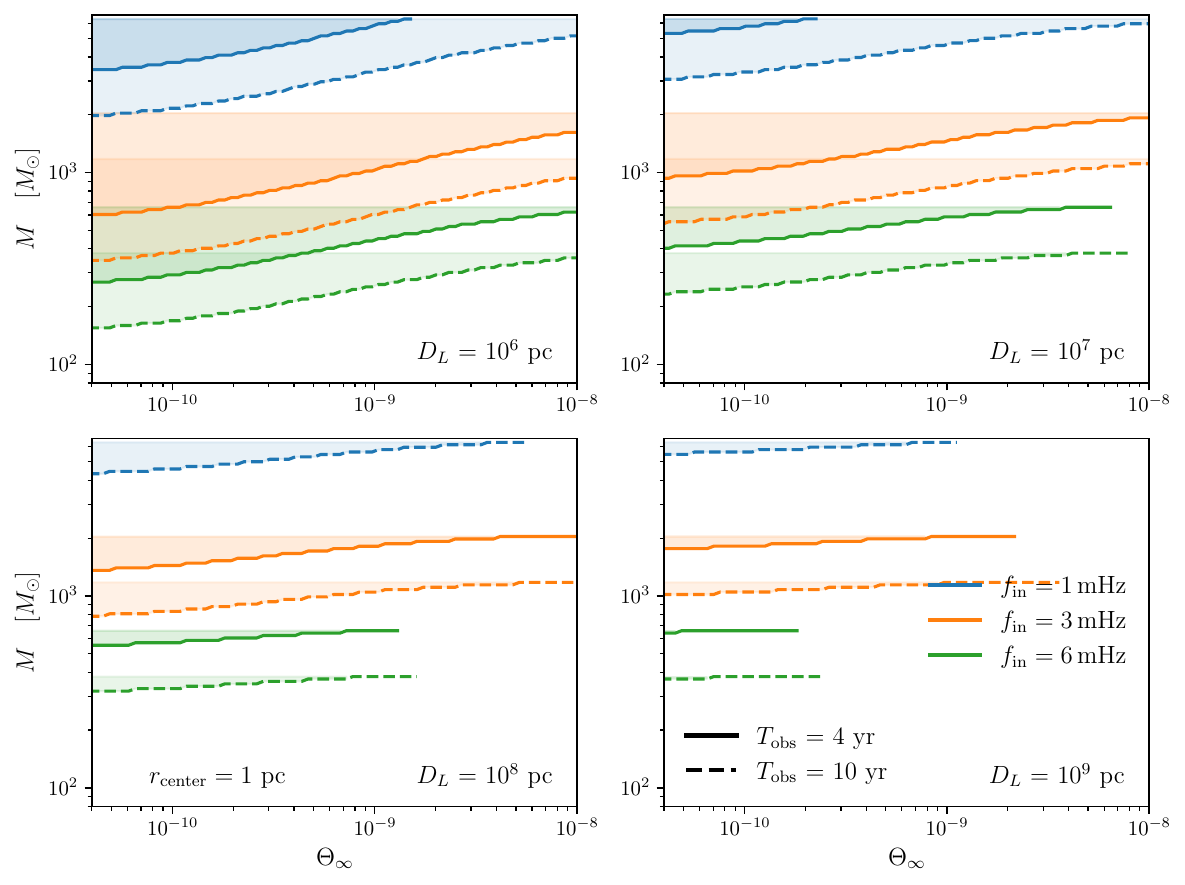}
     \includegraphics[width=0.85\linewidth]{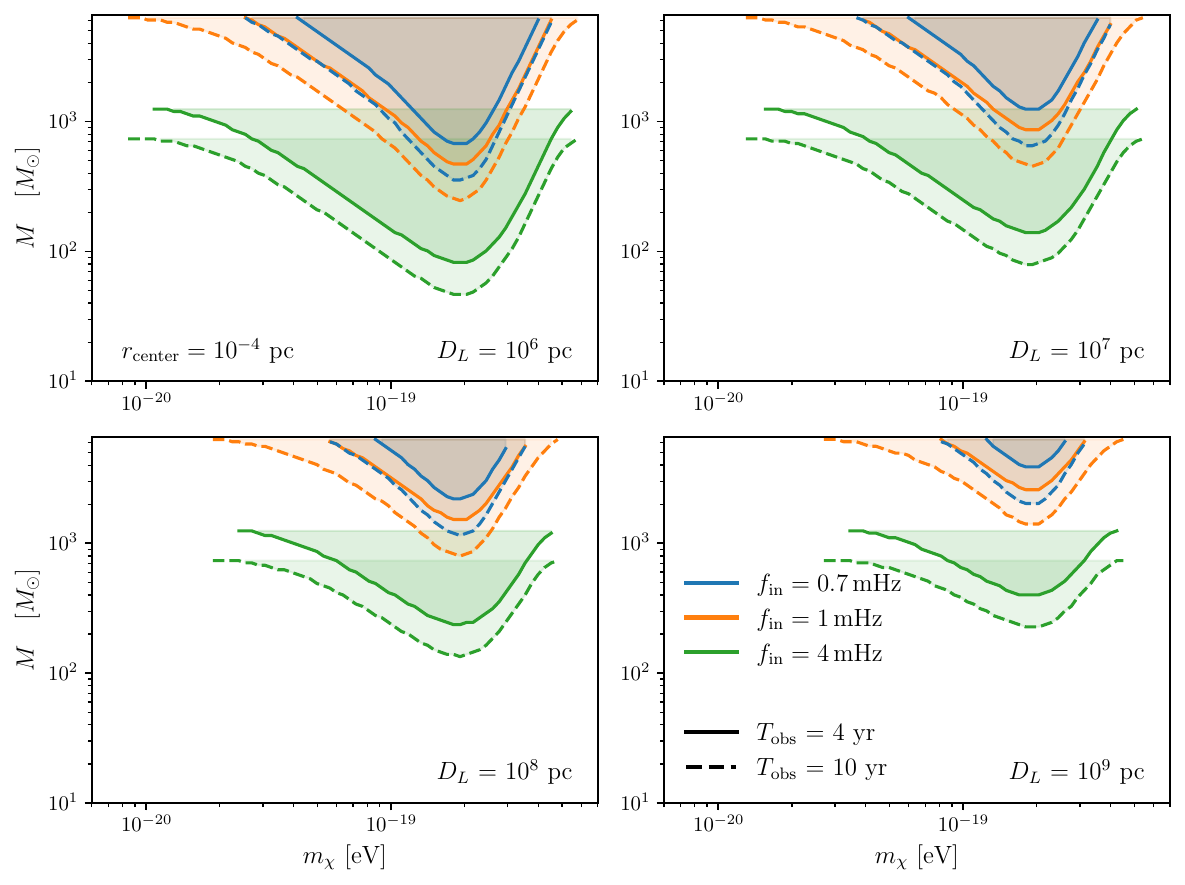}
     % \vspace{-10pt}
    \caption{Same as Fig.~\ref{fig: extragalactic detectability} but with $q = 0.1$.}
    \label{fig: extragalactic detectability with q = 0.1}
\end{figure*}

Includes Fig.~\ref{fig: galactic detectability with q = 0.1} and \ref{fig: extragalactic detectability with q = 0.1}.

\bibliography{refs}% Produces the bibliography via BibTeX.

%apsrev4-2.bst 2019-01-14 (MD) hand-edited version of apsrev4-1.bst
%Control: key (0)
%Control: author (8) initials jnrlst
%Control: editor formatted (1) identically to author
%Control: production of article title (0) allowed
%Control: page (0) single
%Control: year (1) truncated
%Control: production of eprint (0) enabled
\providecommand{\noopsort}[1]{}\providecommand{\singleletter}[1]{#1}%
\begin{thebibliography}{74}%
\makeatletter
\providecommand \@ifxundefined [1]{%
 \@ifx{#1\undefined}
}%
\providecommand \@ifnum [1]{%
 \ifnum #1\expandafter \@firstoftwo
 \else \expandafter \@secondoftwo
 \fi
}%
\providecommand \@ifx [1]{%
 \ifx #1\expandafter \@firstoftwo
 \else \expandafter \@secondoftwo
 \fi
}%
\providecommand \natexlab [1]{#1}%
\providecommand \enquote  [1]{``#1''}%
\providecommand \bibnamefont  [1]{#1}%
\providecommand \bibfnamefont [1]{#1}%
\providecommand \citenamefont [1]{#1}%
\providecommand \href@noop [0]{\@secondoftwo}%
\providecommand \href [0]{\begingroup \@sanitize@url \@href}%
\providecommand \@href[1]{\@@startlink{#1}\@@href}%
\providecommand \@@href[1]{\endgroup#1\@@endlink}%
\providecommand \@sanitize@url [0]{\catcode `\\12\catcode `\$12\catcode `\&12\catcode `\#12\catcode `\^12\catcode `\_12\catcode `\%12\relax}%
\providecommand \@@startlink[1]{}%
\providecommand \@@endlink[0]{}%
\providecommand \url  [0]{\begingroup\@sanitize@url \@url }%
\providecommand \@url [1]{\endgroup\@href {#1}{\urlprefix }}%
\providecommand \urlprefix  [0]{URL }%
\providecommand \Eprint [0]{\href }%
\providecommand \doibase [0]{https://doi.org/}%
\providecommand \selectlanguage [0]{\@gobble}%
\providecommand \bibinfo  [0]{\@secondoftwo}%
\providecommand \bibfield  [0]{\@secondoftwo}%
\providecommand \translation [1]{[#1]}%
\providecommand \BibitemOpen [0]{}%
\providecommand \bibitemStop [0]{}%
\providecommand \bibitemNoStop [0]{.\EOS\space}%
\providecommand \EOS [0]{\spacefactor3000\relax}%
\providecommand \BibitemShut  [1]{\csname bibitem#1\endcsname}%
\let\auto@bib@innerbib\@empty
%</preamble>
\bibitem [{\citenamefont {Eda}\ \emph {et~al.}(2013)\citenamefont {Eda}, \citenamefont {Itoh}, \citenamefont {Kuroyanagi},\ and\ \citenamefont {Silk}}]{Eda:2013gg}%
  \BibitemOpen
  \bibfield  {author} {\bibinfo {author} {\bibfnamefont {K.}~\bibnamefont {Eda}}, \bibinfo {author} {\bibfnamefont {Y.}~\bibnamefont {Itoh}}, \bibinfo {author} {\bibfnamefont {S.}~\bibnamefont {Kuroyanagi}},\ and\ \bibinfo {author} {\bibfnamefont {J.}~\bibnamefont {Silk}},\ }\bibfield  {title} {\bibinfo {title} {{New Probe of Dark-Matter Properties: Gravitational Waves from an Intermediate-Mass Black Hole Embedded in a Dark-Matter Minispike}},\ }\href {https://doi.org/10.1103/PhysRevLett.110.221101} {\bibfield  {journal} {\bibinfo  {journal} {Phys. Rev. Lett.}\ }\textbf {\bibinfo {volume} {110}},\ \bibinfo {pages} {221101} (\bibinfo {year} {2013})},\ \Eprint {https://arxiv.org/abs/1301.5971} {arXiv:1301.5971 [gr-qc]} \BibitemShut {NoStop}%
\bibitem [{\citenamefont {Macedo}\ \emph {et~al.}(2013)\citenamefont {Macedo}, \citenamefont {Pani}, \citenamefont {Cardoso},\ and\ \citenamefont {Crispino}}]{Macedo:2013qea}%
  \BibitemOpen
  \bibfield  {author} {\bibinfo {author} {\bibfnamefont {C.~F.~B.}\ \bibnamefont {Macedo}}, \bibinfo {author} {\bibfnamefont {P.}~\bibnamefont {Pani}}, \bibinfo {author} {\bibfnamefont {V.}~\bibnamefont {Cardoso}},\ and\ \bibinfo {author} {\bibfnamefont {L.~C.~B.}\ \bibnamefont {Crispino}},\ }\bibfield  {title} {\bibinfo {title} {{Into the lair: gravitational-wave signatures of dark matter}},\ }\href {https://doi.org/10.1088/0004-637X/774/1/48} {\bibfield  {journal} {\bibinfo  {journal} {Astrophys. J.}\ }\textbf {\bibinfo {volume} {774}},\ \bibinfo {pages} {48} (\bibinfo {year} {2013})},\ \Eprint {https://arxiv.org/abs/1302.2646} {arXiv:1302.2646 [gr-qc]} \BibitemShut {NoStop}%
\bibitem [{\citenamefont {Kavanagh}\ \emph {et~al.}(2020)\citenamefont {Kavanagh}, \citenamefont {Nichols}, \citenamefont {Bertone},\ and\ \citenamefont {Gaggero}}]{Kavanagh:2020cfn}%
  \BibitemOpen
  \bibfield  {author} {\bibinfo {author} {\bibfnamefont {B.~J.}\ \bibnamefont {Kavanagh}}, \bibinfo {author} {\bibfnamefont {D.~A.}\ \bibnamefont {Nichols}}, \bibinfo {author} {\bibfnamefont {G.}~\bibnamefont {Bertone}},\ and\ \bibinfo {author} {\bibfnamefont {D.}~\bibnamefont {Gaggero}},\ }\bibfield  {title} {\bibinfo {title} {{Detecting dark matter around black holes with gravitational waves: Effects of dark-matter dynamics on the gravitational waveform}},\ }\href {https://doi.org/10.1103/PhysRevD.102.083006} {\bibfield  {journal} {\bibinfo  {journal} {Phys. Rev. D}\ }\textbf {\bibinfo {volume} {102}},\ \bibinfo {pages} {083006} (\bibinfo {year} {2020})},\ \Eprint {https://arxiv.org/abs/2002.12811} {arXiv:2002.12811 [gr-qc]} \BibitemShut {NoStop}%
\bibitem [{\citenamefont {Dai}\ \emph {et~al.}(2023)\citenamefont {Dai}, \citenamefont {Gong}, \citenamefont {Zhao},\ and\ \citenamefont {Jiang}}]{Dai:2023cft}%
  \BibitemOpen
  \bibfield  {author} {\bibinfo {author} {\bibfnamefont {N.}~\bibnamefont {Dai}}, \bibinfo {author} {\bibfnamefont {Y.}~\bibnamefont {Gong}}, \bibinfo {author} {\bibfnamefont {Y.}~\bibnamefont {Zhao}},\ and\ \bibinfo {author} {\bibfnamefont {T.}~\bibnamefont {Jiang}},\ }\bibfield  {title} {\bibinfo {title} {{Extreme mass ratio inspirals in galaxies with dark matter halos}},\ }\href@noop {} {\  (\bibinfo {year} {2023})},\ \Eprint {https://arxiv.org/abs/2301.05088} {arXiv:2301.05088 [gr-qc]} \BibitemShut {NoStop}%
\bibitem [{\citenamefont {Traykova}\ \emph {et~al.}(2023)\citenamefont {Traykova}, \citenamefont {Vicente}, \citenamefont {Clough}, \citenamefont {Helfer}, \citenamefont {Berti}, \citenamefont {Ferreira},\ and\ \citenamefont {Hui}}]{Traykova:2023qyv}%
  \BibitemOpen
  \bibfield  {author} {\bibinfo {author} {\bibfnamefont {D.}~\bibnamefont {Traykova}}, \bibinfo {author} {\bibfnamefont {R.}~\bibnamefont {Vicente}}, \bibinfo {author} {\bibfnamefont {K.}~\bibnamefont {Clough}}, \bibinfo {author} {\bibfnamefont {T.}~\bibnamefont {Helfer}}, \bibinfo {author} {\bibfnamefont {E.}~\bibnamefont {Berti}}, \bibinfo {author} {\bibfnamefont {P.~G.}\ \bibnamefont {Ferreira}},\ and\ \bibinfo {author} {\bibfnamefont {L.}~\bibnamefont {Hui}},\ }\bibfield  {title} {\bibinfo {title} {{Relativistic drag forces on black holes from scalar dark matter clouds of all sizes}},\ }\href {https://doi.org/10.1103/PhysRevD.108.L121502} {\bibfield  {journal} {\bibinfo  {journal} {Phys. Rev. D}\ }\textbf {\bibinfo {volume} {108}},\ \bibinfo {pages} {L121502} (\bibinfo {year} {2023})},\ \Eprint {https://arxiv.org/abs/2305.10492} {arXiv:2305.10492 [gr-qc]} \BibitemShut {NoStop}%
\bibitem [{\citenamefont {Akil}\ and\ \citenamefont {Ding}(2023)}]{Akil:2023kym}%
  \BibitemOpen
  \bibfield  {author} {\bibinfo {author} {\bibfnamefont {A.}~\bibnamefont {Akil}}\ and\ \bibinfo {author} {\bibfnamefont {Q.}~\bibnamefont {Ding}},\ }\bibfield  {title} {\bibinfo {title} {{A dark matter probe in accreting pulsar-black hole binaries}},\ }\href {https://doi.org/10.1088/1475-7516/2023/09/011} {\bibfield  {journal} {\bibinfo  {journal} {JCAP}\ }\textbf {\bibinfo {volume} {09}},\ \bibinfo {pages} {011}},\ \Eprint {https://arxiv.org/abs/2304.08824} {arXiv:2304.08824 [astro-ph.HE]} \BibitemShut {NoStop}%
\bibitem [{\citenamefont {Kadota}\ \emph {et~al.}(2024)\citenamefont {Kadota}, \citenamefont {Kim}, \citenamefont {Ko},\ and\ \citenamefont {Yang}}]{Kadota:2023wlm}%
  \BibitemOpen
  \bibfield  {author} {\bibinfo {author} {\bibfnamefont {K.}~\bibnamefont {Kadota}}, \bibinfo {author} {\bibfnamefont {J.~H.}\ \bibnamefont {Kim}}, \bibinfo {author} {\bibfnamefont {P.}~\bibnamefont {Ko}},\ and\ \bibinfo {author} {\bibfnamefont {X.-Y.}\ \bibnamefont {Yang}},\ }\bibfield  {title} {\bibinfo {title} {{Gravitational wave probes on self-interacting dark matter surrounding an intermediate mass black hole}},\ }\href {https://doi.org/10.1103/PhysRevD.109.015022} {\bibfield  {journal} {\bibinfo  {journal} {Phys. Rev. D}\ }\textbf {\bibinfo {volume} {109}},\ \bibinfo {pages} {015022} (\bibinfo {year} {2024})},\ \Eprint {https://arxiv.org/abs/2306.10828} {arXiv:2306.10828 [hep-ph]} \BibitemShut {NoStop}%
\bibitem [{\citenamefont {Mitra}\ \emph {et~al.}(2024)\citenamefont {Mitra}, \citenamefont {Chakraborty}, \citenamefont {Vicente},\ and\ \citenamefont {Feng}}]{Mitra:2023sny}%
  \BibitemOpen
  \bibfield  {author} {\bibinfo {author} {\bibfnamefont {S.}~\bibnamefont {Mitra}}, \bibinfo {author} {\bibfnamefont {S.}~\bibnamefont {Chakraborty}}, \bibinfo {author} {\bibfnamefont {R.}~\bibnamefont {Vicente}},\ and\ \bibinfo {author} {\bibfnamefont {J.~C.}\ \bibnamefont {Feng}},\ }\bibfield  {title} {\bibinfo {title} {{Probing the quantum nature of black holes with ultralight boson environments}},\ }\href {https://doi.org/10.1103/PhysRevD.110.084012} {\bibfield  {journal} {\bibinfo  {journal} {Phys. Rev. D}\ }\textbf {\bibinfo {volume} {110}},\ \bibinfo {pages} {084012} (\bibinfo {year} {2024})},\ \Eprint {https://arxiv.org/abs/2312.06783} {arXiv:2312.06783 [gr-qc]} \BibitemShut {NoStop}%
\bibitem [{\citenamefont {Cai}\ \emph {et~al.}(2024)\citenamefont {Cai}, \citenamefont {Dom\`enech}, \citenamefont {Ganz}, \citenamefont {Jiang}, \citenamefont {Lin},\ and\ \citenamefont {Wang}}]{Cai:2023ykr}%
  \BibitemOpen
  \bibfield  {author} {\bibinfo {author} {\bibfnamefont {Y.-F.}\ \bibnamefont {Cai}}, \bibinfo {author} {\bibfnamefont {G.}~\bibnamefont {Dom\`enech}}, \bibinfo {author} {\bibfnamefont {A.}~\bibnamefont {Ganz}}, \bibinfo {author} {\bibfnamefont {J.}~\bibnamefont {Jiang}}, \bibinfo {author} {\bibfnamefont {C.}~\bibnamefont {Lin}},\ and\ \bibinfo {author} {\bibfnamefont {B.}~\bibnamefont {Wang}},\ }\bibfield  {title} {\bibinfo {title} {{Parametric resonance of gravitational waves in general scalar-tensor theories}},\ }\href {https://doi.org/10.1088/1475-7516/2024/10/027} {\bibfield  {journal} {\bibinfo  {journal} {JCAP}\ }\textbf {\bibinfo {volume} {10}},\ \bibinfo {pages} {027}},\ \Eprint {https://arxiv.org/abs/2311.18546} {arXiv:2311.18546 [gr-qc]} \BibitemShut {NoStop}%
\bibitem [{\citenamefont {Califano}\ \emph {et~al.}(2024)\citenamefont {Califano}, \citenamefont {De~Martino},\ and\ \citenamefont {Lazkoz}}]{Califano:2024xzt}%
  \BibitemOpen
  \bibfield  {author} {\bibinfo {author} {\bibfnamefont {M.}~\bibnamefont {Califano}}, \bibinfo {author} {\bibfnamefont {I.}~\bibnamefont {De~Martino}},\ and\ \bibinfo {author} {\bibfnamefont {R.}~\bibnamefont {Lazkoz}},\ }\bibfield  {title} {\bibinfo {title} {{Probing interacting dark sector with the next generation of gravitational-wave detectors}},\ }\href {https://doi.org/10.1103/PhysRevD.110.083519} {\bibfield  {journal} {\bibinfo  {journal} {Phys. Rev. D}\ }\textbf {\bibinfo {volume} {110}},\ \bibinfo {pages} {083519} (\bibinfo {year} {2024})},\ \Eprint {https://arxiv.org/abs/2410.06152} {arXiv:2410.06152 [gr-qc]} \BibitemShut {NoStop}%
\bibitem [{\citenamefont {Bramante}\ \emph {et~al.}(2020)\citenamefont {Bramante}, \citenamefont {Buchanan}, \citenamefont {Goodman},\ and\ \citenamefont {Lodhi}}]{Bramante:2019fhi}%
  \BibitemOpen
  \bibfield  {author} {\bibinfo {author} {\bibfnamefont {J.}~\bibnamefont {Bramante}}, \bibinfo {author} {\bibfnamefont {A.}~\bibnamefont {Buchanan}}, \bibinfo {author} {\bibfnamefont {A.}~\bibnamefont {Goodman}},\ and\ \bibinfo {author} {\bibfnamefont {E.}~\bibnamefont {Lodhi}},\ }\bibfield  {title} {\bibinfo {title} {{Terrestrial and Martian Heat Flow Limits on Dark Matter}},\ }\href {https://doi.org/10.1103/PhysRevD.101.043001} {\bibfield  {journal} {\bibinfo  {journal} {Phys. Rev. D}\ }\textbf {\bibinfo {volume} {101}},\ \bibinfo {pages} {043001} (\bibinfo {year} {2020})},\ \Eprint {https://arxiv.org/abs/1909.11683} {arXiv:1909.11683 [hep-ph]} \BibitemShut {NoStop}%
\bibitem [{\citenamefont {{Mack}}\ \emph {et~al.}(2007)\citenamefont {{Mack}}, \citenamefont {{Beacom}},\ and\ \citenamefont {{Bertone}}}]{Mack:Beacom:2007}%
  \BibitemOpen
  \bibfield  {author} {\bibinfo {author} {\bibfnamefont {G.~D.}\ \bibnamefont {{Mack}}}, \bibinfo {author} {\bibfnamefont {J.~F.}\ \bibnamefont {{Beacom}}},\ and\ \bibinfo {author} {\bibfnamefont {G.}~\bibnamefont {{Bertone}}},\ }\bibfield  {title} {\bibinfo {title} {{Towards closing the window on strongly interacting dark matter: Far-reaching constraints from Earth's heat flow}},\ }\href {https://doi.org/10.1103/PhysRevD.76.043523} {\bibfield  {journal} {\bibinfo  {journal} {\prd}\ }\textbf {\bibinfo {volume} {76}},\ \bibinfo {eid} {043523} (\bibinfo {year} {2007})},\ \Eprint {https://arxiv.org/abs/0705.4298} {arXiv:0705.4298 [astro-ph]} \BibitemShut {NoStop}%
\bibitem [{\citenamefont {Leane}\ and\ \citenamefont {Smirnov}(2021)}]{leane2021exoplanets}%
  \BibitemOpen
  \bibfield  {author} {\bibinfo {author} {\bibfnamefont {R.~K.}\ \bibnamefont {Leane}}\ and\ \bibinfo {author} {\bibfnamefont {J.}~\bibnamefont {Smirnov}},\ }\bibfield  {title} {\bibinfo {title} {{Exoplanets as Sub-GeV Dark Matter Detectors}},\ }\href {https://doi.org/10.1103/PhysRevLett.126.161101} {\bibfield  {journal} {\bibinfo  {journal} {Phys. Rev. Lett.}\ }\textbf {\bibinfo {volume} {126}},\ \bibinfo {pages} {161101} (\bibinfo {year} {2021})},\ \Eprint {https://arxiv.org/abs/2010.00015} {arXiv:2010.00015 [hep-ph]} \BibitemShut {NoStop}%
\bibitem [{\citenamefont {Bell}\ and\ \citenamefont {Petraki}(2011)}]{Bell:2011sn}%
  \BibitemOpen
  \bibfield  {author} {\bibinfo {author} {\bibfnamefont {N.~F.}\ \bibnamefont {Bell}}\ and\ \bibinfo {author} {\bibfnamefont {K.}~\bibnamefont {Petraki}},\ }\bibfield  {title} {\bibinfo {title} {{Enhanced neutrino signals from dark matter annihilation in the Sun via metastable mediators}},\ }\href {https://doi.org/10.1088/1475-7516/2011/04/003} {\bibfield  {journal} {\bibinfo  {journal} {JCAP}\ }\textbf {\bibinfo {volume} {04}},\ \bibinfo {pages} {003}},\ \Eprint {https://arxiv.org/abs/1102.2958} {arXiv:1102.2958 [hep-ph]} \BibitemShut {NoStop}%
\bibitem [{\citenamefont {Navas}(2020)}]{km3netsun}%
  \BibitemOpen
  \bibfield  {author} {\bibinfo {author} {\bibfnamefont {S.}~\bibnamefont {Navas}} (\bibinfo {collaboration} {KM3NeT}),\ }\bibfield  {title} {\bibinfo {title} {{Dark Matter Searches from the Sun with the KM3NeT-ORCA detector}},\ }\href@noop {} {\bibfield  {journal} {\bibinfo  {journal} {PoS}\ }\textbf {\bibinfo {volume} {ICRC2019}},\ \bibinfo {pages} {536} (\bibinfo {year} {2020})}\BibitemShut {NoStop}%
\bibitem [{\citenamefont {In}\ and\ \citenamefont {Wiebe}(2018)}]{In:2017kcf}%
  \BibitemOpen
  \bibfield  {author} {\bibinfo {author} {\bibfnamefont {S.}~\bibnamefont {In}}\ and\ \bibinfo {author} {\bibfnamefont {K.}~\bibnamefont {Wiebe}} (\bibinfo {collaboration} {IceCube}),\ }\bibfield  {title} {\bibinfo {title} {{Latest results and sensitivities for solar dark matter searches with IceCube}},\ }\href {https://doi.org/10.22323/1.301.0912} {\bibfield  {journal} {\bibinfo  {journal} {PoS}\ }\textbf {\bibinfo {volume} {ICRC2017}},\ \bibinfo {pages} {912} (\bibinfo {year} {2018})}\BibitemShut {NoStop}%
\bibitem [{\citenamefont {Gould}(1987)}]{Gould:1987ju}%
  \BibitemOpen
  \bibfield  {author} {\bibinfo {author} {\bibfnamefont {A.}~\bibnamefont {Gould}},\ }\bibfield  {title} {\bibinfo {title} {{{WIMP} Distribution in and Evaporation From the Sun}},\ }\href {https://doi.org/10.1086/165652} {\bibfield  {journal} {\bibinfo  {journal} {Astrophys. J.}\ }\textbf {\bibinfo {volume} {321}},\ \bibinfo {pages} {560} (\bibinfo {year} {1987})}\BibitemShut {NoStop}%
%%CITATION = ASJOA,321,560;%%
\bibitem [{\citenamefont {Press}\ and\ \citenamefont {Spergel}(1985)}]{Press:1985ug}%
  \BibitemOpen
  \bibfield  {author} {\bibinfo {author} {\bibfnamefont {W.~H.}\ \bibnamefont {Press}}\ and\ \bibinfo {author} {\bibfnamefont {D.~N.}\ \bibnamefont {Spergel}},\ }\bibfield  {title} {\bibinfo {title} {{Capture by the sun of a galactic population of weakly interacting massive particles}},\ }\href {https://doi.org/10.1086/163485} {\bibfield  {journal} {\bibinfo  {journal} {Astrophys. J.}\ }\textbf {\bibinfo {volume} {296}},\ \bibinfo {pages} {679} (\bibinfo {year} {1985})}\BibitemShut {NoStop}%
\bibitem [{\citenamefont {Peter}(2009)}]{Peter:2009mk}%
  \BibitemOpen
  \bibfield  {author} {\bibinfo {author} {\bibfnamefont {A.~H.}\ \bibnamefont {Peter}},\ }\bibfield  {title} {\bibinfo {title} {{Dark matter in the solar system II: WIMP annihilation rates in the Sun}},\ }\href {https://doi.org/10.1103/PhysRevD.79.103532} {\bibfield  {journal} {\bibinfo  {journal} {Phys. Rev. D}\ }\textbf {\bibinfo {volume} {79}},\ \bibinfo {pages} {103532} (\bibinfo {year} {2009})},\ \Eprint {https://arxiv.org/abs/0902.1347} {arXiv:0902.1347 [astro-ph.HE]} \BibitemShut {NoStop}%
\bibitem [{\citenamefont {Garani}\ and\ \citenamefont {Palomares-Ruiz}(2017)}]{Garani:2017jcj}%
  \BibitemOpen
  \bibfield  {author} {\bibinfo {author} {\bibfnamefont {R.}~\bibnamefont {Garani}}\ and\ \bibinfo {author} {\bibfnamefont {S.}~\bibnamefont {Palomares-Ruiz}},\ }\bibfield  {title} {\bibinfo {title} {{Dark matter in the Sun: scattering off electrons vs nucleons}},\ }\href {https://doi.org/10.1088/1475-7516/2017/05/007} {\bibfield  {journal} {\bibinfo  {journal} {JCAP}\ }\textbf {\bibinfo {volume} {05}},\ \bibinfo {pages} {007}},\ \Eprint {https://arxiv.org/abs/1702.02768} {arXiv:1702.02768 [hep-ph]} \BibitemShut {NoStop}%
\bibitem [{\citenamefont {Kouvaris}(2015)}]{Kouvaris:2015nsa}%
  \BibitemOpen
  \bibfield  {author} {\bibinfo {author} {\bibfnamefont {C.}~\bibnamefont {Kouvaris}},\ }\bibfield  {title} {\bibinfo {title} {{Probing Light Dark Matter via Evaporation from the Sun}},\ }\href {https://doi.org/10.1103/PhysRevD.92.075001} {\bibfield  {journal} {\bibinfo  {journal} {Phys. Rev. D}\ }\textbf {\bibinfo {volume} {92}},\ \bibinfo {pages} {075001} (\bibinfo {year} {2015})},\ \Eprint {https://arxiv.org/abs/1506.04316} {arXiv:1506.04316 [hep-ph]} \BibitemShut {NoStop}%
\bibitem [{\citenamefont {{Croon}}\ \emph {et~al.}(2024)\citenamefont {{Croon}}, \citenamefont {{Sakstein}}, \citenamefont {{Smirnov}},\ and\ \citenamefont {{Streeter}}}]{Croon:2024}%
  \BibitemOpen
  \bibfield  {author} {\bibinfo {author} {\bibfnamefont {D.}~\bibnamefont {{Croon}}}, \bibinfo {author} {\bibfnamefont {J.}~\bibnamefont {{Sakstein}}}, \bibinfo {author} {\bibfnamefont {J.}~\bibnamefont {{Smirnov}}},\ and\ \bibinfo {author} {\bibfnamefont {J.}~\bibnamefont {{Streeter}}},\ }\bibfield  {title} {\bibinfo {title} {{Dark Dwarfs: Dark Matter-Powered Sub-Stellar Objects Awaiting Discovery at the Galactic Center}},\ }\href {https://doi.org/10.48550/arXiv.2408.00822} {\bibfield  {journal} {\bibinfo  {journal} {arXiv e-prints}\ ,\ \bibinfo {eid} {arXiv:2408.00822}} (\bibinfo {year} {2024})},\ \Eprint {https://arxiv.org/abs/2408.00822} {arXiv:2408.00822 [hep-ph]} \BibitemShut {NoStop}%
\bibitem [{\citenamefont {Leung}\ \emph {et~al.}(2013)\citenamefont {Leung}, \citenamefont {Chu}, \citenamefont {Lin},\ and\ \citenamefont {Wong}}]{Leung:2013pra}%
  \BibitemOpen
  \bibfield  {author} {\bibinfo {author} {\bibfnamefont {S.~C.}\ \bibnamefont {Leung}}, \bibinfo {author} {\bibfnamefont {M.~C.}\ \bibnamefont {Chu}}, \bibinfo {author} {\bibfnamefont {L.~M.}\ \bibnamefont {Lin}},\ and\ \bibinfo {author} {\bibfnamefont {K.~W.}\ \bibnamefont {Wong}},\ }\bibfield  {title} {\bibinfo {title} {{Dark-matter admixed white dwarfs}},\ }\href {https://doi.org/10.1103/PhysRevD.87.123506} {\bibfield  {journal} {\bibinfo  {journal} {Phys. Rev.}\ }\textbf {\bibinfo {volume} {D87}},\ \bibinfo {pages} {123506} (\bibinfo {year} {2013})},\ \Eprint {https://arxiv.org/abs/1305.6142} {arXiv:1305.6142 [astro-ph.CO]} \BibitemShut {NoStop}%
%%CITATION = ARXIV:1305.6142;%%
\bibitem [{\citenamefont {Graham}\ \emph {et~al.}(2018{\natexlab{a}})\citenamefont {Graham}, \citenamefont {Janish}, \citenamefont {Narayan}, \citenamefont {Rajendran},\ and\ \citenamefont {Riggins}}]{Graham:2018efk}%
  \BibitemOpen
  \bibfield  {author} {\bibinfo {author} {\bibfnamefont {P.~W.}\ \bibnamefont {Graham}}, \bibinfo {author} {\bibfnamefont {R.}~\bibnamefont {Janish}}, \bibinfo {author} {\bibfnamefont {V.}~\bibnamefont {Narayan}}, \bibinfo {author} {\bibfnamefont {S.}~\bibnamefont {Rajendran}},\ and\ \bibinfo {author} {\bibfnamefont {P.}~\bibnamefont {Riggins}},\ }\bibfield  {title} {\bibinfo {title} {{White Dwarfs as Dark Matter Detectors}},\ }\href {https://doi.org/10.1103/PhysRevD.98.115027} {\bibfield  {journal} {\bibinfo  {journal} {Phys. Rev.}\ }\textbf {\bibinfo {volume} {D98}},\ \bibinfo {pages} {115027} (\bibinfo {year} {2018}{\natexlab{a}})},\ \Eprint {https://arxiv.org/abs/1805.07381} {arXiv:1805.07381 [hep-ph]} \BibitemShut {NoStop}%
%%CITATION = ARXIV:1805.07381;%%
\bibitem [{\citenamefont {Acevedo}\ and\ \citenamefont {Bramante}(2019)}]{Acevedo:2019gre}%
  \BibitemOpen
  \bibfield  {author} {\bibinfo {author} {\bibfnamefont {J.~F.}\ \bibnamefont {Acevedo}}\ and\ \bibinfo {author} {\bibfnamefont {J.}~\bibnamefont {Bramante}},\ }\bibfield  {title} {\bibinfo {title} {{Supernovae Sparked By Dark Matter in White Dwarfs}},\ }\href {https://doi.org/10.1103/PhysRevD.100.043020} {\bibfield  {journal} {\bibinfo  {journal} {Phys. Rev.}\ }\textbf {\bibinfo {volume} {D100}},\ \bibinfo {pages} {043020} (\bibinfo {year} {2019})},\ \Eprint {https://arxiv.org/abs/1904.11993} {arXiv:1904.11993 [hep-ph]} \BibitemShut {NoStop}%
%%CITATION = ARXIV:1904.11993;%%
\bibitem [{\citenamefont {Graham}\ \emph {et~al.}(2018{\natexlab{b}})\citenamefont {Graham}, \citenamefont {Janish}, \citenamefont {Narayan}, \citenamefont {Rajendran},\ and\ \citenamefont {Riggins}}]{graham2018white}%
  \BibitemOpen
  \bibfield  {author} {\bibinfo {author} {\bibfnamefont {P.~W.}\ \bibnamefont {Graham}}, \bibinfo {author} {\bibfnamefont {R.}~\bibnamefont {Janish}}, \bibinfo {author} {\bibfnamefont {V.}~\bibnamefont {Narayan}}, \bibinfo {author} {\bibfnamefont {S.}~\bibnamefont {Rajendran}},\ and\ \bibinfo {author} {\bibfnamefont {P.}~\bibnamefont {Riggins}},\ }\bibfield  {title} {\bibinfo {title} {{White Dwarfs as Dark Matter Detectors}},\ }\href {https://doi.org/10.1103/PhysRevD.98.115027} {\bibfield  {journal} {\bibinfo  {journal} {Phys. Rev. D}\ }\textbf {\bibinfo {volume} {98}},\ \bibinfo {pages} {115027} (\bibinfo {year} {2018}{\natexlab{b}})},\ \Eprint {https://arxiv.org/abs/1805.07381} {arXiv:1805.07381 [hep-ph]} \BibitemShut {NoStop}%
\bibitem [{\citenamefont {Goldman}\ and\ \citenamefont {Nussinov}(1989)}]{Goldman:1989}%
  \BibitemOpen
  \bibfield  {author} {\bibinfo {author} {\bibfnamefont {I.}~\bibnamefont {Goldman}}\ and\ \bibinfo {author} {\bibfnamefont {S.}~\bibnamefont {Nussinov}},\ }\bibfield  {title} {\bibinfo {title} {Weakly interacting massive particles and neutron stars},\ }\href {https://doi.org/10.1103/PhysRevD.40.3221} {\bibfield  {journal} {\bibinfo  {journal} {Phys. Rev. D}\ }\textbf {\bibinfo {volume} {40}},\ \bibinfo {pages} {3221} (\bibinfo {year} {1989})}\BibitemShut {NoStop}%
\bibitem [{\citenamefont {{Gould}}\ \emph {et~al.}(1990)\citenamefont {{Gould}}, \citenamefont {{Draine}}, \citenamefont {{Romani}},\ and\ \citenamefont {{Nussinov}}}]{Gould:1990}%
  \BibitemOpen
  \bibfield  {author} {\bibinfo {author} {\bibfnamefont {A.}~\bibnamefont {{Gould}}}, \bibinfo {author} {\bibfnamefont {B.~T.}\ \bibnamefont {{Draine}}}, \bibinfo {author} {\bibfnamefont {R.~W.}\ \bibnamefont {{Romani}}},\ and\ \bibinfo {author} {\bibfnamefont {S.}~\bibnamefont {{Nussinov}}},\ }\bibfield  {title} {\bibinfo {title} {{Neutron stars: Graveyard of charged dark matter}},\ }\href {https://doi.org/10.1016/0370-2693(90)91745-W} {\bibfield  {journal} {\bibinfo  {journal} {Physics Letters B}\ }\textbf {\bibinfo {volume} {238}},\ \bibinfo {pages} {337} (\bibinfo {year} {1990})}\BibitemShut {NoStop}%
\bibitem [{\citenamefont {Kouvaris}(2008)}]{Kouvaris:2007ay}%
  \BibitemOpen
  \bibfield  {author} {\bibinfo {author} {\bibfnamefont {C.}~\bibnamefont {Kouvaris}},\ }\bibfield  {title} {\bibinfo {title} {{WIMP Annihilation and Cooling of Neutron Stars}},\ }\href {https://doi.org/10.1103/PhysRevD.77.023006} {\bibfield  {journal} {\bibinfo  {journal} {Phys. Rev. D}\ }\textbf {\bibinfo {volume} {77}},\ \bibinfo {pages} {023006} (\bibinfo {year} {2008})},\ \Eprint {https://arxiv.org/abs/0708.2362} {arXiv:0708.2362 [astro-ph]} \BibitemShut {NoStop}%
\bibitem [{\citenamefont {Garani}\ and\ \citenamefont {Heeck}(2019)}]{Garani:2019fpa}%
  \BibitemOpen
  \bibfield  {author} {\bibinfo {author} {\bibfnamefont {R.}~\bibnamefont {Garani}}\ and\ \bibinfo {author} {\bibfnamefont {J.}~\bibnamefont {Heeck}},\ }\bibfield  {title} {\bibinfo {title} {{Dark matter interactions with muons in neutron stars}},\ }\href {https://doi.org/10.1103/PhysRevD.100.035039} {\bibfield  {journal} {\bibinfo  {journal} {Phys. Rev.}\ }\textbf {\bibinfo {volume} {D100}},\ \bibinfo {pages} {035039} (\bibinfo {year} {2019})},\ \Eprint {https://arxiv.org/abs/1906.10145} {arXiv:1906.10145 [hep-ph]} \BibitemShut {NoStop}%
%%CITATION = ARXIV:1906.10145;%%
\bibitem [{\citenamefont {Kouvaris}\ and\ \citenamefont {Tinyakov}(2010)}]{Kouvaris_2010}%
  \BibitemOpen
  \bibfield  {author} {\bibinfo {author} {\bibfnamefont {C.}~\bibnamefont {Kouvaris}}\ and\ \bibinfo {author} {\bibfnamefont {P.}~\bibnamefont {Tinyakov}},\ }\bibfield  {title} {\bibinfo {title} {Can neutron stars constrain dark matter?},\ }\bibfield  {journal} {\bibinfo  {journal} {Physical Review D}\ }\textbf {\bibinfo {volume} {82}},\ \href {https://doi.org/10.1103/physrevd.82.063531} {10.1103/physrevd.82.063531} (\bibinfo {year} {2010})\BibitemShut {NoStop}%
\bibitem [{\citenamefont {Bramante}\ and\ \citenamefont {Linden}(2014)}]{Bramante:2014zca}%
  \BibitemOpen
  \bibfield  {author} {\bibinfo {author} {\bibfnamefont {J.}~\bibnamefont {Bramante}}\ and\ \bibinfo {author} {\bibfnamefont {T.}~\bibnamefont {Linden}},\ }\bibfield  {title} {\bibinfo {title} {{Detecting Dark Matter with Imploding Pulsars in the Galactic Center}},\ }\href {https://doi.org/10.1103/PhysRevLett.113.191301} {\bibfield  {journal} {\bibinfo  {journal} {Phys. Rev. Lett.}\ }\textbf {\bibinfo {volume} {113}},\ \bibinfo {pages} {191301} (\bibinfo {year} {2014})},\ \Eprint {https://arxiv.org/abs/1405.1031} {arXiv:1405.1031 [astro-ph.HE]} \BibitemShut {NoStop}%
\bibitem [{\citenamefont {Gresham}\ and\ \citenamefont {Zurek}(2018)}]{Gresham:2018rqo}%
  \BibitemOpen
  \bibfield  {author} {\bibinfo {author} {\bibfnamefont {M.~I.}\ \bibnamefont {Gresham}}\ and\ \bibinfo {author} {\bibfnamefont {K.~M.}\ \bibnamefont {Zurek}},\ }\bibfield  {title} {\bibinfo {title} {{Asymmetric Dark Stars and Neutron Star Stability}},\ }\href@noop {} {\  (\bibinfo {year} {2018})},\ \Eprint {https://arxiv.org/abs/1809.08254} {arXiv:1809.08254 [astro-ph.CO]} \BibitemShut {NoStop}%
%%CITATION = ARXIV:1809.08254;%%
\bibitem [{\citenamefont {Bramante}\ \emph {et~al.}(2018)\citenamefont {Bramante}, \citenamefont {Linden},\ and\ \citenamefont {Tsai}}]{Bramante:2017ulk}%
  \BibitemOpen
  \bibfield  {author} {\bibinfo {author} {\bibfnamefont {J.}~\bibnamefont {Bramante}}, \bibinfo {author} {\bibfnamefont {T.}~\bibnamefont {Linden}},\ and\ \bibinfo {author} {\bibfnamefont {Y.-D.}\ \bibnamefont {Tsai}},\ }\bibfield  {title} {\bibinfo {title} {{Searching for dark matter with neutron star mergers and quiet kilonovae}},\ }\href {https://doi.org/10.1103/PhysRevD.97.055016} {\bibfield  {journal} {\bibinfo  {journal} {Phys. Rev.}\ }\textbf {\bibinfo {volume} {D97}},\ \bibinfo {pages} {055016} (\bibinfo {year} {2018})},\ \Eprint {https://arxiv.org/abs/1706.00001} {arXiv:1706.00001 [hep-ph]} \BibitemShut {NoStop}%
%%CITATION = ARXIV:1706.00001;%%
\bibitem [{\citenamefont {de~Lavallaz}\ and\ \citenamefont {Fairbairn}(2010)}]{de_neutron_2010}%
  \BibitemOpen
  \bibfield  {author} {\bibinfo {author} {\bibfnamefont {A.}~\bibnamefont {de~Lavallaz}}\ and\ \bibinfo {author} {\bibfnamefont {M.}~\bibnamefont {Fairbairn}},\ }\bibfield  {title} {\bibinfo {title} {{Neutron Stars as Dark Matter Probes}},\ }\href {https://doi.org/10.1103/PhysRevD.81.123521} {\bibfield  {journal} {\bibinfo  {journal} {Phys. Rev. D}\ }\textbf {\bibinfo {volume} {81}},\ \bibinfo {pages} {123521} (\bibinfo {year} {2010})},\ \Eprint {https://arxiv.org/abs/1004.0629} {arXiv:1004.0629 [astro-ph.GA]} \BibitemShut {NoStop}%
\bibitem [{\citenamefont {{Abbott}}\ \emph {et~al.}(2021)\citenamefont {{Abbott}} \emph {et~al.}}]{Abbott_O3a_2021}%
  \BibitemOpen
  \bibfield  {author} {\bibinfo {author} {\bibfnamefont {R.}~\bibnamefont {{Abbott}}} \emph {et~al.},\ }\bibfield  {title} {\bibinfo {title} {{GWTC-2: Compact Binary Coalescences Observed by LIGO and Virgo during the First Half of the Third Observing Run}},\ }\href {https://doi.org/10.1103/PhysRevX.11.021053} {\bibfield  {journal} {\bibinfo  {journal} {Physical Review X}\ }\textbf {\bibinfo {volume} {11}},\ \bibinfo {eid} {021053} (\bibinfo {year} {2021})},\ \Eprint {https://arxiv.org/abs/2010.14527} {arXiv:2010.14527 [gr-qc]} \BibitemShut {NoStop}%
\bibitem [{\citenamefont {{Abbott}}\ \emph {et~al.}(2023)\citenamefont {{Abbott}} \emph {et~al.}}]{Abbott_O3b_2021}%
  \BibitemOpen
  \bibfield  {author} {\bibinfo {author} {\bibfnamefont {R.}~\bibnamefont {{Abbott}}} \emph {et~al.},\ }\bibfield  {title} {\bibinfo {title} {{GWTC-3: Compact Binary Coalescences Observed by LIGO and Virgo during the Second Part of the Third Observing Run}},\ }\href {https://doi.org/10.1103/PhysRevX.13.041039} {\bibfield  {journal} {\bibinfo  {journal} {Physical Review X}\ }\textbf {\bibinfo {volume} {13}},\ \bibinfo {eid} {041039} (\bibinfo {year} {2023})},\ \Eprint {https://arxiv.org/abs/2111.03606} {arXiv:2111.03606 [gr-qc]} \BibitemShut {NoStop}%
\bibitem [{\citenamefont {Seoane}\ \emph {et~al.}(2023)\citenamefont {Seoane} \emph {et~al.}}]{LISA:2022yao}%
  \BibitemOpen
  \bibfield  {author} {\bibinfo {author} {\bibfnamefont {P.~A.}\ \bibnamefont {Seoane}} \emph {et~al.} (\bibinfo {collaboration} {LISA}),\ }\bibfield  {title} {\bibinfo {title} {{Astrophysics with the Laser Interferometer Space Antenna}},\ }\href {https://doi.org/10.1007/s41114-022-00041-y} {\bibfield  {journal} {\bibinfo  {journal} {Living Rev. Rel.}\ }\textbf {\bibinfo {volume} {26}},\ \bibinfo {pages} {2} (\bibinfo {year} {2023})},\ \Eprint {https://arxiv.org/abs/2203.06016} {arXiv:2203.06016 [gr-qc]} \BibitemShut {NoStop}%
\bibitem [{\citenamefont {{Robson}}\ \emph {et~al.}(2019)\citenamefont {{Robson}}, \citenamefont {{Cornish}},\ and\ \citenamefont {{Liu}}}]{Robson:Cornish:2019}%
  \BibitemOpen
  \bibfield  {author} {\bibinfo {author} {\bibfnamefont {T.}~\bibnamefont {{Robson}}}, \bibinfo {author} {\bibfnamefont {N.~J.}\ \bibnamefont {{Cornish}}},\ and\ \bibinfo {author} {\bibfnamefont {C.}~\bibnamefont {{Liu}}},\ }\bibfield  {title} {\bibinfo {title} {{The construction and use of LISA sensitivity curves}},\ }\href {https://doi.org/10.1088/1361-6382/ab1101} {\bibfield  {journal} {\bibinfo  {journal} {Classical and Quantum Gravity}\ }\textbf {\bibinfo {volume} {36}},\ \bibinfo {eid} {105011} (\bibinfo {year} {2019})},\ \Eprint {https://arxiv.org/abs/1803.01944} {arXiv:1803.01944 [astro-ph.HE]} \BibitemShut {NoStop}%
\bibitem [{\citenamefont {Wagg}\ \emph {et~al.}(2022)\citenamefont {Wagg}, \citenamefont {Broekgaarden}, \citenamefont {de~Mink}, \citenamefont {van Son}, \citenamefont {Frankel},\ and\ \citenamefont {Justham}}]{Wagg:2021cst}%
  \BibitemOpen
  \bibfield  {author} {\bibinfo {author} {\bibfnamefont {T.}~\bibnamefont {Wagg}}, \bibinfo {author} {\bibfnamefont {F.~S.}\ \bibnamefont {Broekgaarden}}, \bibinfo {author} {\bibfnamefont {S.~E.}\ \bibnamefont {de~Mink}}, \bibinfo {author} {\bibfnamefont {L.~A.~C.}\ \bibnamefont {van Son}}, \bibinfo {author} {\bibfnamefont {N.}~\bibnamefont {Frankel}},\ and\ \bibinfo {author} {\bibfnamefont {S.}~\bibnamefont {Justham}},\ }\bibfield  {title} {\bibinfo {title} {{Gravitational Wave Sources in Our Galactic Backyard: Predictions for BHBH, BHNS, and NSNS Binaries Detectable with LISA}},\ }\href {https://doi.org/10.3847/1538-4357/ac8675} {\bibfield  {journal} {\bibinfo  {journal} {Astrophys. J.}\ }\textbf {\bibinfo {volume} {937}},\ \bibinfo {pages} {118} (\bibinfo {year} {2022})},\ \Eprint {https://arxiv.org/abs/2111.13704} {arXiv:2111.13704 [astro-ph.HE]} \BibitemShut {NoStop}%
\bibitem [{\citenamefont {Jani}\ \emph {et~al.}(2019)\citenamefont {Jani}, \citenamefont {Shoemaker},\ and\ \citenamefont {Cutler}}]{Jani:2019ffg}%
  \BibitemOpen
  \bibfield  {author} {\bibinfo {author} {\bibfnamefont {K.}~\bibnamefont {Jani}}, \bibinfo {author} {\bibfnamefont {D.}~\bibnamefont {Shoemaker}},\ and\ \bibinfo {author} {\bibfnamefont {C.}~\bibnamefont {Cutler}},\ }\bibfield  {title} {\bibinfo {title} {{Detectability of Intermediate-Mass Black Holes in Multiband Gravitational Wave Astronomy}},\ }\href {https://doi.org/10.1038/s41550-019-0932-7} {\bibfield  {journal} {\bibinfo  {journal} {Nature Astron.}\ }\textbf {\bibinfo {volume} {4}},\ \bibinfo {pages} {260} (\bibinfo {year} {2019})},\ \Eprint {https://arxiv.org/abs/1908.04985} {arXiv:1908.04985 [gr-qc]} \BibitemShut {NoStop}%
\bibitem [{\citenamefont {{Schive}}\ \emph {et~al.}(2014)\citenamefont {{Schive}}, \citenamefont {{Chiueh}},\ and\ \citenamefont {{Broadhurst}}}]{Schive:2014}%
  \BibitemOpen
  \bibfield  {author} {\bibinfo {author} {\bibfnamefont {H.-Y.}\ \bibnamefont {{Schive}}}, \bibinfo {author} {\bibfnamefont {T.}~\bibnamefont {{Chiueh}}},\ and\ \bibinfo {author} {\bibfnamefont {T.}~\bibnamefont {{Broadhurst}}},\ }\bibfield  {title} {\bibinfo {title} {{Cosmic structure as the quantum interference of a coherent dark wave}},\ }\href {https://doi.org/10.1038/nphys2996} {\bibfield  {journal} {\bibinfo  {journal} {Nature Physics}\ }\textbf {\bibinfo {volume} {10}},\ \bibinfo {pages} {496} (\bibinfo {year} {2014})},\ \Eprint {https://arxiv.org/abs/1406.6586} {arXiv:1406.6586 [astro-ph.GA]} \BibitemShut {NoStop}%
\bibitem [{\citenamefont {Klypin}\ \emph {et~al.}(1999)\citenamefont {Klypin}, \citenamefont {Kravtsov}, \citenamefont {Valenzuela},\ and\ \citenamefont {Prada}}]{Klypin:1999uc}%
  \BibitemOpen
  \bibfield  {author} {\bibinfo {author} {\bibfnamefont {A.~A.}\ \bibnamefont {Klypin}}, \bibinfo {author} {\bibfnamefont {A.~V.}\ \bibnamefont {Kravtsov}}, \bibinfo {author} {\bibfnamefont {O.}~\bibnamefont {Valenzuela}},\ and\ \bibinfo {author} {\bibfnamefont {F.}~\bibnamefont {Prada}},\ }\bibfield  {title} {\bibinfo {title} {{Where are the missing Galactic satellites?}},\ }\href {https://doi.org/10.1086/307643} {\bibfield  {journal} {\bibinfo  {journal} {Astrophys. J.}\ }\textbf {\bibinfo {volume} {522}},\ \bibinfo {pages} {82} (\bibinfo {year} {1999})},\ \Eprint {https://arxiv.org/abs/astro-ph/9901240} {arXiv:astro-ph/9901240} \BibitemShut {NoStop}%
\bibitem [{\citenamefont {{de Blok}}(2010)}]{deBlok:2010}%
  \BibitemOpen
  \bibfield  {author} {\bibinfo {author} {\bibfnamefont {W.~J.~G.}\ \bibnamefont {{de Blok}}},\ }\bibfield  {title} {\bibinfo {title} {{The Core-Cusp Problem}},\ }\href {https://doi.org/10.1155/2010/789293} {\bibfield  {journal} {\bibinfo  {journal} {Advances in Astronomy}\ }\textbf {\bibinfo {volume} {2010}},\ \bibinfo {eid} {789293} (\bibinfo {year} {2010})},\ \Eprint {https://arxiv.org/abs/0910.3538} {arXiv:0910.3538 [astro-ph.CO]} \BibitemShut {NoStop}%
\bibitem [{\citenamefont {Hoyle}\ and\ \citenamefont {Lyttleton}(1941)}]{hoyle1941accretion}%
  \BibitemOpen
  \bibfield  {author} {\bibinfo {author} {\bibfnamefont {F.}~\bibnamefont {Hoyle}}\ and\ \bibinfo {author} {\bibfnamefont {R.}~\bibnamefont {Lyttleton}},\ }\bibfield  {title} {\bibinfo {title} {On the accretion theory of stellar evolution},\ }\href@noop {} {\bibfield  {journal} {\bibinfo  {journal} {Monthly Notices of the Royal Astronomical Society, Vol. 101, p. 227}\ }\textbf {\bibinfo {volume} {101}},\ \bibinfo {pages} {227} (\bibinfo {year} {1941})}\BibitemShut {NoStop}%
\bibitem [{\citenamefont {{Bondi}}(1952)}]{Bondi1952}%
  \BibitemOpen
  \bibfield  {author} {\bibinfo {author} {\bibfnamefont {H.}~\bibnamefont {{Bondi}}},\ }\bibfield  {title} {\bibinfo {title} {{On spherically symmetrical accretion}},\ }\href {https://doi.org/10.1093/mnras/112.2.195} {\bibfield  {journal} {\bibinfo  {journal} {\mnras}\ }\textbf {\bibinfo {volume} {112}},\ \bibinfo {pages} {195} (\bibinfo {year} {1952})}\BibitemShut {NoStop}%
\bibitem [{\citenamefont {Michel}(1972)}]{michel1972accretion}%
  \BibitemOpen
  \bibfield  {author} {\bibinfo {author} {\bibfnamefont {F.~C.}\ \bibnamefont {Michel}},\ }\bibfield  {title} {\bibinfo {title} {Accretion of matter by condensed objects},\ }\href@noop {} {\bibfield  {journal} {\bibinfo  {journal} {Astrophysics and Space Science}\ }\textbf {\bibinfo {volume} {15}},\ \bibinfo {pages} {153} (\bibinfo {year} {1972})}\BibitemShut {NoStop}%
\bibitem [{\citenamefont {{Tejeda}}\ \emph {et~al.}(2020)\citenamefont {{Tejeda}}, \citenamefont {{Aguayo-Ortiz}},\ and\ \citenamefont {{Hernandez}}}]{Tejeda:2020}%
  \BibitemOpen
  \bibfield  {author} {\bibinfo {author} {\bibfnamefont {E.}~\bibnamefont {{Tejeda}}}, \bibinfo {author} {\bibfnamefont {A.}~\bibnamefont {{Aguayo-Ortiz}}},\ and\ \bibinfo {author} {\bibfnamefont {X.}~\bibnamefont {{Hernandez}}},\ }\bibfield  {title} {\bibinfo {title} {{Choked Accretion onto a Schwarzschild Black Hole: A Hydrodynamical Jet-launching Mechanism}},\ }\href {https://doi.org/10.3847/1538-4357/ab7ffe} {\bibfield  {journal} {\bibinfo  {journal} {\apj}\ }\textbf {\bibinfo {volume} {893}},\ \bibinfo {eid} {81} (\bibinfo {year} {2020})},\ \Eprint {https://arxiv.org/abs/1909.01527} {arXiv:1909.01527 [astro-ph.HE]} \BibitemShut {NoStop}%
\bibitem [{\citenamefont {Aguayo-Ortiz}\ \emph {et~al.}(2021)\citenamefont {Aguayo-Ortiz}, \citenamefont {Tejeda}, \citenamefont {Sarbach},\ and\ \citenamefont {L{\'o}pez-C{\'a}mara}}]{aguayo2021spherical}%
  \BibitemOpen
  \bibfield  {author} {\bibinfo {author} {\bibfnamefont {A.}~\bibnamefont {Aguayo-Ortiz}}, \bibinfo {author} {\bibfnamefont {E.}~\bibnamefont {Tejeda}}, \bibinfo {author} {\bibfnamefont {O.}~\bibnamefont {Sarbach}},\ and\ \bibinfo {author} {\bibfnamefont {D.}~\bibnamefont {L{\'o}pez-C{\'a}mara}},\ }\bibfield  {title} {\bibinfo {title} {Spherical accretion: Bondi, michel, and rotating black holes},\ }\href@noop {} {\bibfield  {journal} {\bibinfo  {journal} {Monthly Notices of the Royal Astronomical Society}\ }\textbf {\bibinfo {volume} {504}},\ \bibinfo {pages} {5039} (\bibinfo {year} {2021})}\BibitemShut {NoStop}%
\bibitem [{\citenamefont {Avelino}\ and\ \citenamefont {Ferreira}(2015)}]{Avelino_DM_soundspeed}%
  \BibitemOpen
  \bibfield  {author} {\bibinfo {author} {\bibfnamefont {P.}~\bibnamefont {Avelino}}\ and\ \bibinfo {author} {\bibfnamefont {V.}~\bibnamefont {Ferreira}},\ }\bibfield  {title} {\bibinfo {title} {Constraints on the dark matter sound speed from galactic scales: The cases of the modified and extended chaplygin gas},\ }\href {https://doi.org/10.1103/PhysRevD.91.083508} {\bibfield  {journal} {\bibinfo  {journal} {PRD}\ }\textbf {\bibinfo {volume} {91}},\ \bibinfo {pages} {083508} (\bibinfo {year} {2015})}\BibitemShut {NoStop}%
\bibitem [{\citenamefont {{Bondi}}\ and\ \citenamefont {{Hoyle}}(1944)}]{Bond_Hoyle_1944}%
  \BibitemOpen
  \bibfield  {author} {\bibinfo {author} {\bibfnamefont {H.}~\bibnamefont {{Bondi}}}\ and\ \bibinfo {author} {\bibfnamefont {F.}~\bibnamefont {{Hoyle}}},\ }\bibfield  {title} {\bibinfo {title} {{On the mechanism of accretion by stars}},\ }\href {https://doi.org/10.1093/mnras/104.5.273} {\bibfield  {journal} {\bibinfo  {journal} {\mnras}\ }\textbf {\bibinfo {volume} {104}},\ \bibinfo {pages} {273} (\bibinfo {year} {1944})}\BibitemShut {NoStop}%
\bibitem [{\citenamefont {Gondolo}\ and\ \citenamefont {Silk}(1999)}]{Gondolo:1999ef}%
  \BibitemOpen
  \bibfield  {author} {\bibinfo {author} {\bibfnamefont {P.}~\bibnamefont {Gondolo}}\ and\ \bibinfo {author} {\bibfnamefont {J.}~\bibnamefont {Silk}},\ }\bibfield  {title} {\bibinfo {title} {{Dark matter annihilation at the galactic center}},\ }\href {https://doi.org/10.1103/PhysRevLett.83.1719} {\bibfield  {journal} {\bibinfo  {journal} {Phys. Rev. Lett.}\ }\textbf {\bibinfo {volume} {83}},\ \bibinfo {pages} {1719} (\bibinfo {year} {1999})},\ \Eprint {https://arxiv.org/abs/astro-ph/9906391} {arXiv:astro-ph/9906391} \BibitemShut {NoStop}%
\bibitem [{\citenamefont {Davies}\ and\ \citenamefont {Mocz}(2020)}]{Davies_2020}%
  \BibitemOpen
  \bibfield  {author} {\bibinfo {author} {\bibfnamefont {E.~Y.}\ \bibnamefont {Davies}}\ and\ \bibinfo {author} {\bibfnamefont {P.}~\bibnamefont {Mocz}},\ }\bibfield  {title} {\bibinfo {title} {Fuzzy dark matter soliton cores around supermassive black holes},\ }\href {https://doi.org/10.1093/mnras/staa202} {\bibfield  {journal} {\bibinfo  {journal} {Monthly Notices of the Royal Astronomical Society}\ }\textbf {\bibinfo {volume} {492}},\ \bibinfo {pages} {5721–5729} (\bibinfo {year} {2020})}\BibitemShut {NoStop}%
\bibitem [{\citenamefont {Kim}\ \emph {et~al.}(2023)\citenamefont {Kim}, \citenamefont {Lenoci}, \citenamefont {Stomberg},\ and\ \citenamefont {Xue}}]{Kim:2022mdj}%
  \BibitemOpen
  \bibfield  {author} {\bibinfo {author} {\bibfnamefont {H.}~\bibnamefont {Kim}}, \bibinfo {author} {\bibfnamefont {A.}~\bibnamefont {Lenoci}}, \bibinfo {author} {\bibfnamefont {I.}~\bibnamefont {Stomberg}},\ and\ \bibinfo {author} {\bibfnamefont {X.}~\bibnamefont {Xue}},\ }\bibfield  {title} {\bibinfo {title} {{Adiabatically compressed wave dark matter halo and intermediate-mass-ratio inspirals}},\ }\href {https://doi.org/10.1103/PhysRevD.107.083005} {\bibfield  {journal} {\bibinfo  {journal} {Phys. Rev. D}\ }\textbf {\bibinfo {volume} {107}},\ \bibinfo {pages} {083005} (\bibinfo {year} {2023})},\ \Eprint {https://arxiv.org/abs/2212.07528} {arXiv:2212.07528 [astro-ph.GA]} \BibitemShut {NoStop}%
\bibitem [{\citenamefont {{Unruh}}(1976)}]{Unruh_1976}%
  \BibitemOpen
  \bibfield  {author} {\bibinfo {author} {\bibfnamefont {W.~G.}\ \bibnamefont {{Unruh}}},\ }\bibfield  {title} {\bibinfo {title} {{Absorption cross section of small black holes}},\ }\href {https://doi.org/10.1103/PhysRevD.14.3251} {\bibfield  {journal} {\bibinfo  {journal} {\prd}\ }\textbf {\bibinfo {volume} {14}},\ \bibinfo {pages} {3251} (\bibinfo {year} {1976})}\BibitemShut {NoStop}%
\bibitem [{\citenamefont {Hui}\ \emph {et~al.}(2017)\citenamefont {Hui}, \citenamefont {Ostriker}, \citenamefont {Tremaine},\ and\ \citenamefont {Witten}}]{hui2017ultralight}%
  \BibitemOpen
  \bibfield  {author} {\bibinfo {author} {\bibfnamefont {L.}~\bibnamefont {Hui}}, \bibinfo {author} {\bibfnamefont {J.~P.}\ \bibnamefont {Ostriker}}, \bibinfo {author} {\bibfnamefont {S.}~\bibnamefont {Tremaine}},\ and\ \bibinfo {author} {\bibfnamefont {E.}~\bibnamefont {Witten}},\ }\bibfield  {title} {\bibinfo {title} {Ultralight scalars as cosmological dark matter},\ }\href@noop {} {\bibfield  {journal} {\bibinfo  {journal} {Physical Review D}\ }\textbf {\bibinfo {volume} {95}},\ \bibinfo {pages} {043541} (\bibinfo {year} {2017})}\BibitemShut {NoStop}%
\bibitem [{\citenamefont {De~Martino}\ \emph {et~al.}(2020)\citenamefont {De~Martino}, \citenamefont {Broadhurst}, \citenamefont {Tye}, \citenamefont {Chiueh},\ and\ \citenamefont {Schive}}]{de2020dynamical}%
  \BibitemOpen
  \bibfield  {author} {\bibinfo {author} {\bibfnamefont {I.}~\bibnamefont {De~Martino}}, \bibinfo {author} {\bibfnamefont {T.}~\bibnamefont {Broadhurst}}, \bibinfo {author} {\bibfnamefont {S.-H.~H.}\ \bibnamefont {Tye}}, \bibinfo {author} {\bibfnamefont {T.}~\bibnamefont {Chiueh}},\ and\ \bibinfo {author} {\bibfnamefont {H.-Y.}\ \bibnamefont {Schive}},\ }\bibfield  {title} {\bibinfo {title} {Dynamical evidence of a dark solitonic core of 109$m_\odot$ in the milky way},\ }\href@noop {} {\bibfield  {journal} {\bibinfo  {journal} {Physics of the Dark Universe}\ }\textbf {\bibinfo {volume} {28}},\ \bibinfo {pages} {100503} (\bibinfo {year} {2020})}\BibitemShut {NoStop}%
\bibitem [{\citenamefont {{Farris}}\ \emph {et~al.}(2010)\citenamefont {{Farris}}, \citenamefont {{Liu}},\ and\ \citenamefont {{Shapiro}}}]{Farris:2010}%
  \BibitemOpen
  \bibfield  {author} {\bibinfo {author} {\bibfnamefont {B.~D.}\ \bibnamefont {{Farris}}}, \bibinfo {author} {\bibfnamefont {Y.~T.}\ \bibnamefont {{Liu}}},\ and\ \bibinfo {author} {\bibfnamefont {S.~L.}\ \bibnamefont {{Shapiro}}},\ }\bibfield  {title} {\bibinfo {title} {{Binary black hole mergers in gaseous environments: ``Binary Bondi`` and ``binary Bondi-Hoyle-Lyttleton'' accretion}},\ }\href {https://doi.org/10.1103/PhysRevD.81.084008} {\bibfield  {journal} {\bibinfo  {journal} {\prd}\ }\textbf {\bibinfo {volume} {81}},\ \bibinfo {eid} {084008} (\bibinfo {year} {2010})},\ \Eprint {https://arxiv.org/abs/0912.2096} {arXiv:0912.2096 [astro-ph.HE]} \BibitemShut {NoStop}%
\bibitem [{\citenamefont {{Kaaz}}\ \emph {et~al.}(2019)\citenamefont {{Kaaz}}, \citenamefont {{Antoni}},\ and\ \citenamefont {{Ramirez-Ruiz}}}]{Kaaz:2019}%
  \BibitemOpen
  \bibfield  {author} {\bibinfo {author} {\bibfnamefont {N.}~\bibnamefont {{Kaaz}}}, \bibinfo {author} {\bibfnamefont {A.}~\bibnamefont {{Antoni}}},\ and\ \bibinfo {author} {\bibfnamefont {E.}~\bibnamefont {{Ramirez-Ruiz}}},\ }\bibfield  {title} {\bibinfo {title} {{Bondi-Hoyle-Lyttleton Accretion onto Star Clusters}},\ }\href {https://doi.org/10.3847/1538-4357/ab158b} {\bibfield  {journal} {\bibinfo  {journal} {\apj}\ }\textbf {\bibinfo {volume} {876}},\ \bibinfo {eid} {142} (\bibinfo {year} {2019})},\ \Eprint {https://arxiv.org/abs/1901.03649} {arXiv:1901.03649 [astro-ph.HE]} \BibitemShut {NoStop}%
\bibitem [{\citenamefont {{Comerford}}\ \emph {et~al.}(2019)\citenamefont {{Comerford}}, \citenamefont {{Izzard}}, \citenamefont {{Booth}},\ and\ \citenamefont {{Rosotti}}}]{Comerford:2019}%
  \BibitemOpen
  \bibfield  {author} {\bibinfo {author} {\bibfnamefont {T.~A.~F.}\ \bibnamefont {{Comerford}}}, \bibinfo {author} {\bibfnamefont {R.~G.}\ \bibnamefont {{Izzard}}}, \bibinfo {author} {\bibfnamefont {R.~A.}\ \bibnamefont {{Booth}}},\ and\ \bibinfo {author} {\bibfnamefont {G.}~\bibnamefont {{Rosotti}}},\ }\bibfield  {title} {\bibinfo {title} {{Bondi-Hoyle-Lyttleton accretion by binary stars}},\ }\href {https://doi.org/10.1093/mnras/stz2977} {\bibfield  {journal} {\bibinfo  {journal} {\mnras}\ }\textbf {\bibinfo {volume} {490}},\ \bibinfo {pages} {5196} (\bibinfo {year} {2019})},\ \Eprint {https://arxiv.org/abs/1910.13353} {arXiv:1910.13353 [astro-ph.SR]} \BibitemShut {NoStop}%
\bibitem [{\citenamefont {Peters}(1964)}]{peters1964gravitational}%
  \BibitemOpen
  \bibfield  {author} {\bibinfo {author} {\bibfnamefont {P.~C.}\ \bibnamefont {Peters}},\ }\bibfield  {title} {\bibinfo {title} {Gravitational radiation and the motion of two point masses},\ }\href@noop {} {\bibfield  {journal} {\bibinfo  {journal} {Physical Review}\ }\textbf {\bibinfo {volume} {136}},\ \bibinfo {pages} {B1224} (\bibinfo {year} {1964})}\BibitemShut {NoStop}%
\bibitem [{\citenamefont {Hadjidemetriou}(1963)}]{hadjidemetriou1963two}%
  \BibitemOpen
  \bibfield  {author} {\bibinfo {author} {\bibfnamefont {J.~D.}\ \bibnamefont {Hadjidemetriou}},\ }\bibfield  {title} {\bibinfo {title} {Two-body problem with variable mass: a new approach},\ }\href@noop {} {\bibfield  {journal} {\bibinfo  {journal} {Icarus}\ }\textbf {\bibinfo {volume} {2}},\ \bibinfo {pages} {440} (\bibinfo {year} {1963})}\BibitemShut {NoStop}%
\bibitem [{\citenamefont {{Takahashi}}\ and\ \citenamefont {{Seto}}(2002)}]{Takahashi:Seto:2002}%
  \BibitemOpen
  \bibfield  {author} {\bibinfo {author} {\bibfnamefont {R.}~\bibnamefont {{Takahashi}}}\ and\ \bibinfo {author} {\bibfnamefont {N.}~\bibnamefont {{Seto}}},\ }\bibfield  {title} {\bibinfo {title} {{Parameter Estimation for Galactic Binaries by the Laser Interferometer Space Antenna}},\ }\href {https://doi.org/10.1086/341483} {\bibfield  {journal} {\bibinfo  {journal} {\apj}\ }\textbf {\bibinfo {volume} {575}},\ \bibinfo {pages} {1030} (\bibinfo {year} {2002})},\ \Eprint {https://arxiv.org/abs/astro-ph/0204487} {arXiv:astro-ph/0204487 [astro-ph]} \BibitemShut {NoStop}%
\bibitem [{\citenamefont {Nelemans}\ \emph {et~al.}(2001)\citenamefont {Nelemans}, \citenamefont {Yungelson},\ and\ \citenamefont {Portegies~Zwart}}]{Nelemans:2001hp}%
  \BibitemOpen
  \bibfield  {author} {\bibinfo {author} {\bibfnamefont {G.}~\bibnamefont {Nelemans}}, \bibinfo {author} {\bibfnamefont {L.~R.}\ \bibnamefont {Yungelson}},\ and\ \bibinfo {author} {\bibfnamefont {S.~F.}\ \bibnamefont {Portegies~Zwart}},\ }\bibfield  {title} {\bibinfo {title} {{The gravitational wave signal from the galactic disk population of binaries containing two compact objects}},\ }\href {https://doi.org/10.1051/0004-6361:20010683} {\bibfield  {journal} {\bibinfo  {journal} {Astron. Astrophys.}\ }\textbf {\bibinfo {volume} {375}},\ \bibinfo {pages} {890} (\bibinfo {year} {2001})},\ \Eprint {https://arxiv.org/abs/astro-ph/0105221} {arXiv:astro-ph/0105221} \BibitemShut {NoStop}%
\bibitem [{\citenamefont {{Peters}}\ and\ \citenamefont {{Mathews}}(1963)}]{Peters:Mathews:1963}%
  \BibitemOpen
  \bibfield  {author} {\bibinfo {author} {\bibfnamefont {P.~C.}\ \bibnamefont {{Peters}}}\ and\ \bibinfo {author} {\bibfnamefont {J.}~\bibnamefont {{Mathews}}},\ }\bibfield  {title} {\bibinfo {title} {{Gravitational Radiation from Point Masses in a Keplerian Orbit}},\ }\href {https://doi.org/10.1103/PhysRev.131.435} {\bibfield  {journal} {\bibinfo  {journal} {Physical Review}\ }\textbf {\bibinfo {volume} {131}},\ \bibinfo {pages} {435} (\bibinfo {year} {1963})}\BibitemShut {NoStop}%
\bibitem [{\citenamefont {{Draine}}(2011)}]{Draine:2011}%
  \BibitemOpen
  \bibfield  {author} {\bibinfo {author} {\bibfnamefont {B.~T.}\ \bibnamefont {{Draine}}},\ }\href@noop {} {\emph {\bibinfo {title} {{Physics of the Interstellar and Intergalactic Medium}}}}\ (\bibinfo {year} {2011})\BibitemShut {NoStop}%
\bibitem [{\citenamefont {Ghodla}(2024)}]{Ghodla:2024gda}%
  \BibitemOpen
  \bibfield  {author} {\bibinfo {author} {\bibfnamefont {S.}~\bibnamefont {Ghodla}},\ }\bibfield  {title} {\bibinfo {title} {{The effect of interstellar medium on LVK\textquoteright{}s black holes}},\ }\href {https://doi.org/10.1093/mnras/stae1545} {\bibfield  {journal} {\bibinfo  {journal} {Mon. Not. Roy. Astron. Soc.}\ }\textbf {\bibinfo {volume} {532}},\ \bibinfo {pages} {439} (\bibinfo {year} {2024})},\ \Eprint {https://arxiv.org/abs/2405.17863} {arXiv:2405.17863 [astro-ph.HE]} \BibitemShut {NoStop}%
\bibitem [{\citenamefont {{Shakura}}\ and\ \citenamefont {{Sunyaev}}(1973)}]{Shakura_Sunyaev_1973}%
  \BibitemOpen
  \bibfield  {author} {\bibinfo {author} {\bibfnamefont {N.~I.}\ \bibnamefont {{Shakura}}}\ and\ \bibinfo {author} {\bibfnamefont {R.~A.}\ \bibnamefont {{Sunyaev}}},\ }\bibfield  {title} {\bibinfo {title} {{Black holes in binary systems. Observational appearance.}},\ }\href@noop {} {\bibfield  {journal} {\bibinfo  {journal} {\aap}\ }\textbf {\bibinfo {volume} {24}},\ \bibinfo {pages} {337} (\bibinfo {year} {1973})}\BibitemShut {NoStop}%
\bibitem [{\citenamefont {{Yuan}}\ and\ \citenamefont {{Narayan}}(2014)}]{Yuan_Narayan_2014}%
  \BibitemOpen
  \bibfield  {author} {\bibinfo {author} {\bibfnamefont {F.}~\bibnamefont {{Yuan}}}\ and\ \bibinfo {author} {\bibfnamefont {R.}~\bibnamefont {{Narayan}}},\ }\bibfield  {title} {\bibinfo {title} {{Hot Accretion Flows Around Black Holes}},\ }\href {https://doi.org/10.1146/annurev-astro-082812-141003} {\bibfield  {journal} {\bibinfo  {journal} {\araa}\ }\textbf {\bibinfo {volume} {52}},\ \bibinfo {pages} {529} (\bibinfo {year} {2014})},\ \Eprint {https://arxiv.org/abs/1401.0586} {arXiv:1401.0586 [astro-ph.HE]} \BibitemShut {NoStop}%
\bibitem [{\citenamefont {Pouliasis}\ \emph {et~al.}(2017)\citenamefont {Pouliasis}, \citenamefont {Di~Matteo},\ and\ \citenamefont {Haywood}}]{Pouliasis_2017}%
  \BibitemOpen
  \bibfield  {author} {\bibinfo {author} {\bibfnamefont {E.}~\bibnamefont {Pouliasis}}, \bibinfo {author} {\bibfnamefont {P.}~\bibnamefont {Di~Matteo}},\ and\ \bibinfo {author} {\bibfnamefont {M.}~\bibnamefont {Haywood}},\ }\bibfield  {title} {\bibinfo {title} {A milky way with a massive, centrally concentrated thick disc: new galactic mass models for orbit computations},\ }\href {https://doi.org/10.1051/0004-6361/201527346} {\bibfield  {journal} {\bibinfo  {journal} {Astronomy \&amp; Astrophysics}\ }\textbf {\bibinfo {volume} {598}},\ \bibinfo {pages} {A66} (\bibinfo {year} {2017})}\BibitemShut {NoStop}%
\bibitem [{\citenamefont {{Eilers}}\ \emph {et~al.}(2019)\citenamefont {{Eilers}}, \citenamefont {{Hogg}}, \citenamefont {{Rix}},\ and\ \citenamefont {{Ness}}}]{GAIA:2019}%
  \BibitemOpen
  \bibfield  {author} {\bibinfo {author} {\bibfnamefont {A.-C.}\ \bibnamefont {{Eilers}}}, \bibinfo {author} {\bibfnamefont {D.~W.}\ \bibnamefont {{Hogg}}}, \bibinfo {author} {\bibfnamefont {H.-W.}\ \bibnamefont {{Rix}}},\ and\ \bibinfo {author} {\bibfnamefont {M.~K.}\ \bibnamefont {{Ness}}},\ }\bibfield  {title} {\bibinfo {title} {{The Circular Velocity Curve of the Milky Way from 5 to 25 kpc}},\ }\href {https://doi.org/10.3847/1538-4357/aaf648} {\bibfield  {journal} {\bibinfo  {journal} {\apj}\ }\textbf {\bibinfo {volume} {871}},\ \bibinfo {eid} {120} (\bibinfo {year} {2019})},\ \Eprint {https://arxiv.org/abs/1810.09466} {arXiv:1810.09466 [astro-ph.GA]} \BibitemShut {NoStop}%
\bibitem [{\citenamefont {Reid}\ \emph {et~al.}(2014)\citenamefont {Reid} \emph {et~al.}}]{reid2014trigonometric}%
  \BibitemOpen
  \bibfield  {author} {\bibinfo {author} {\bibfnamefont {M.}~\bibnamefont {Reid}} \emph {et~al.},\ }\bibfield  {title} {\bibinfo {title} {Trigonometric parallaxes of high mass star forming regions: the structure and kinematics of the milky way},\ }\href@noop {} {\bibfield  {journal} {\bibinfo  {journal} {The Astrophysical Journal}\ }\textbf {\bibinfo {volume} {783}},\ \bibinfo {pages} {130} (\bibinfo {year} {2014})}\BibitemShut {NoStop}%
\bibitem [{\citenamefont {Navarro}\ \emph {et~al.}(1997)\citenamefont {Navarro}, \citenamefont {Frenk},\ and\ \citenamefont {White}}]{NFW_1997}%
  \BibitemOpen
  \bibfield  {author} {\bibinfo {author} {\bibfnamefont {J.~F.}\ \bibnamefont {Navarro}}, \bibinfo {author} {\bibfnamefont {C.~S.}\ \bibnamefont {Frenk}},\ and\ \bibinfo {author} {\bibfnamefont {S.~D.}\ \bibnamefont {White}},\ }\bibfield  {title} {\bibinfo {title} {{A Universal Density Profile from Hierarchical Clustering}},\ }\href {https://doi.org/10.1086/304888} {\bibfield  {journal} {\bibinfo  {journal} {\apj}\ }\textbf {\bibinfo {volume} {490}},\ \bibinfo {pages} {493} (\bibinfo {year} {1997})},\ \Eprint {https://arxiv.org/abs/astro-ph/9611107} {arXiv:astro-ph/9611107 [astro-ph]} \BibitemShut {NoStop}%
\bibitem [{\citenamefont {Lin}\ and\ \citenamefont {Li}(2019)}]{lin2019dark}%
  \BibitemOpen
  \bibfield  {author} {\bibinfo {author} {\bibfnamefont {H.-N.}\ \bibnamefont {Lin}}\ and\ \bibinfo {author} {\bibfnamefont {X.}~\bibnamefont {Li}},\ }\bibfield  {title} {\bibinfo {title} {The dark matter profiles in the milky way},\ }\href@noop {} {\bibfield  {journal} {\bibinfo  {journal} {Monthly Notices of the Royal Astronomical Society}\ }\textbf {\bibinfo {volume} {487}},\ \bibinfo {pages} {5679} (\bibinfo {year} {2019})}\BibitemShut {NoStop}%
\end{thebibliography}%

\end{document}